\documentstyle[12pt,epsf]{article} 
\newlength{\dinwidth}
\newlength{\dinmargin}
\newcommand{\be}{\begin{equation}}
\newcommand{\ee}{\end{equation}}
\newcommand{\bes}{\begin{eqnarray}}
\newcommand{\ees}{\end{eqnarray}}
\def\deg{\hbox{$^\circ$}}

\def\gev{\,\hbox{GeV}}

\def\gevv{\,\hbox{GeV}^{2}}

\def\cm{\,\hbox{cm}}

\newcommand{\sgeq} {\raisebox{-.6ex}{${\textstyle\stackrel{>}{\sim}}$}}
\def\epem{$e^+e^-$}
\def\ptm{$<\!\!p_t^{*\,2}\!\!>\,\,$}
\def\gps{$\gamma^* p$ cms}
\begin{document}
\title {\vspace{3cm}
{\bf\Large Inclusive Charged Particle Distributions
  in Deep Inelastic Scattering Events
   at HERA  } \\
\author{\rm ZEUS Collaboration\\ }
}
\date{ }
\maketitle
\vspace{5 cm}
\begin{abstract}
\noindent

A measurement of inclusive charged particle distributions in deep
inelastic $ep$ scattering for $\gamma^* p$ centre-of-mass energies $75< W
< 175$~GeV and momentum transfer squared $10< Q^2 < 160$~GeV$^2$ from the
ZEUS detector at HERA is presented. The differential charged particle
rates in the $\gamma^* p$ centre-of-mass system as a function of the
scaled longitudinal momentum, $x_F$, and of the transverse momentum,
$p_t^*$ and $<\!\!p_t^{*\,2}\!\!>\,\,$ , as a function of $x_F$, $W$ and
$Q^2$ are given. Separate distributions are shown for events with (LRG)
and without (NRG) a rapidity gap with respect to the proton direction. The
data are compared with results from experiments at lower beam energies,
with the naive quark parton model and with parton models including
perturbative QCD corrections. The comparison shows the importance of the
higher order QCD processes. Significant differences of the inclusive
charged particle rates between NRG and LRG events at the same $W$ are
observed. The value of $<\!\!p_t^{*\,2}\!\!>\,\,$ for LRG events with a
hadronic mass $M_X$, which excludes the forward produced baryonic system,
is similar to the $<\!\!p_t^{*\,2}\!\!>\,\,$ value observed in fixed
target experiments at $W \approx M_X$.

\end{abstract}

\vspace{-22cm}
\begin{flushleft}
\tt DESY 95-221 \\
November 1995 \\
\end{flushleft}

\setcounter{page}{0}
\thispagestyle{empty}
\newpage

%
\def\3{\ss}
\textwidth 15.5cm
\parindent 0cm
\footnotesize
\renewcommand{\thepage}{\Roman{page}}
\begin{center}
\begin{large}
The ZEUS Collaboration
\end{large}
\end{center}
M.~Derrick, D.~Krakauer, S.~Magill, D.~Mikunas, B.~Musgrave,
J.~Repond, R.~Stanek, R.L.~Talaga, H.~Zhang \\
{\it Argonne National Laboratory, Argonne, IL, USA}~$^{p}$\\[6pt]
G.~Bari, M.~Basile, L.~Bellagamba, D.~Boscherini, A.~Bruni, G.~Bruni,
P.~Bruni, G.~Cara~Romeo, G.~Castellini$^{1}$,
L.~Cifarelli$^{2}$, F.~Cindolo, A.~Contin, M.~Corradi,
I.~Gialas$^{3}$,
P.~Giusti, G.~Iacobucci, G.~Laurenti, G.~Levi, A.~Margotti,
T.~Massam, R.~Nania, F.~Palmonari, A.~Polini, G.~Sartorelli,\\
Y.~Zamora~Garcia$^{4}$, A.~Zichichi \\
{\it University and INFN Bologna, Bologna, Italy}~$^{f}$
% ...
\\[6pt]
\clearpage
\hspace*{1mm}
$^{ 1}$ also at IROE Florence, Italy  \\
\hspace*{1mm}
$^{ 2}$ now at Univ. of Salerno and INFN Napoli, Italy  \\
\hspace*{1mm}
$^{ 3}$ supported by EU HCM contract ERB-CHRX-CT93-0376 \\
\hspace*{1mm}
$^{ 4}$ supported by Worldlab, Lausanne, Switzerland  \\
\hspace*{1mm}
% ...
$^{21}$ now with OPAL Collaboration, Faculty of Physics at Univ. of
        Freiburg \\
% ...

\begin{tabular}{lp{15cm}}
$^{a}$ & supported by the Natural Sciences and Engineering Research
         Council of Canada (NSERC) \\
$^{b}$ & supported by the FCAR of Qu\'ebec, Canada\\
$^{c}$ & supported by the German Federal Ministry for Education and
         Science, Research and Technology (BMBF), under contract
         numbers 056BN19I, 056FR19P, 056HH19I, 056HH29I, 056SI79I\\
$^{d}$ & supported by the MINERVA Gesellschaft f\"ur Forschung GmbH,
         and by the Israel Academy of Science \\
$^{e}$ & supported by the German Israeli Foundation, and
         by the Israel Academy of Science \\
$^{f}$ & supported by the Italian National Institute for Nuclear Physics
         (INFN) \\
$^{g}$ & supported by the Japanese Ministry of Education, Science and
         Culture (the Monbusho)
         and its grants for Scientific Research\\
$^{h}$ & supported by the Korean Ministry of Education and Korea Science
         and Engineering Foundation \\
$^{i}$ & supported by the Netherlands Foundation for Research on Matter
         (FOM)\\
$^{j}$ & supported by the Polish State Committee for Scientific
         Research, grants No.~115/E-343/SPUB/P03/109/95, 2P03B 244
         08p02, p03, p04 and p05, and the Foundation for Polish-German
         Collaboration (proj. No. 506/92) \\
$^{k}$ & supported by the Polish State Committee for Scientific
         Research (grant No. 2 P03B 083 08) \\
$^{l}$ & partially supported by the German Federal Ministry for
         Education and Science, Research and Technology (BMBF) \\
$^{m}$ & supported by the German Federal Ministry for Education and
         Science, Research and Technology (BMBF), and the Fund of
         Fundamental Research of Russian Ministry of Science and
         Education and by INTAS-Grant No. 93-63 \\
$^{n}$ & supported by the Spanish Ministry of Education and Science
         through funds provided by CICYT \\
$^{o}$ & supported by the Particle Physics and Astronomy Research
         Council \\
$^{p}$ & supported by the US Department of Energy \\
$^{q}$ & supported by the US National Science Foundation
\end{tabular}

\newpage
\pagenumbering{arabic}
\setcounter{page}{1}
\normalsize

\section{ Introduction }

Inclusive particle distributions have been widely studied in deep inelastic
scattering (DIS)~\cite{schmitz93} and \epem~annihilation to investigate the
nature of the quark fragmentation and effects of higher order QCD processes.
The formation of hadrons in DIS is a complicated
process  which cannot be fully calculated in the framework of
perturbative QCD. In order to model this process
it is convenient to distinguish two phases of the hadron formation.
These correspond to a perturbative
phase for QCD processes on the parton level followed by a
non-perturbative fragmentation phase describing the confinement
of the partons to observable hadrons.

In this paper the charged hadron multiplicity distributions are analysed in
the virtual-photon proton centre-of-mass system (\gps), which corresponds to
the centre-of-mass system of the produced hadronic final state
with the invariant mass $W$.
In the naive quark parton model (QPM) the virtual photon hits
a quark in the proton and transfers a four momentum, $q$. The struck
quark and the target remnant system each have an energy of $W/2$ in the
\gps\  and move back-to-back with a `velocity', which corresponds to
a rapidity\footnote{The rapidity is measured with respect to the
virtual photon in the \gps.}
$y_{max}$ proportional to $\pm \ln{W}$. The outgoing quark and
target remnant hadronise into multi-particle final states with limited
$p_t^*$,  where $p_t^*$ is the hadron momentum component transverse
to the virtual photon direction as measured in the \gps.
The width of the rapidity distributions of the produced hadrons  is
proportional to $\ln{W}$, while its height is approximately independent
of $W$. From the measurement of jet profiles in DIS  it is known that the
width of a quark jet is typically two units of rapidity \cite{zeusjet93}.
At high values of $W$, the rapidity range populated
by hadrons can be divided into
three regions: the current jet region from ($y_{max}-2)$ to $y_{max}$,
the region of the target remnant fragmentation from $-y_{max}$ to
($-y_{max}+2)$  
and a plateau region in between.
When analysing hadron distributions as a function of the scaled
longitudinal momentum in the \gps, $x_F$, the current jet region defined
above corresponds to the $x_F$ range $x_F> 0.05$.
If no QCD branching processes on the parton level are considered,
the $x_F$ and $p_t^*$ distributions for $x_F > 0.05$
are predicted to scale in $W$.

In fixed target DIS experiments
\cite{EMC82a,EMCxf,EMC91a} effects of scale-breaking
in the $x_F$ distributions from QCD corrections,
which are expected to soften the observed spectrum with increasing $W$,
are small and could not be unambiguously identified.
On the other hand, the mean square of $p_t^*$, \ptm, has been found to be
very sensitive to higher order QCD effects \cite{EMC80a}.
However, the details of the $p_t^*$ spectra are
also sensitive to non-perturbative fragmentation effects
\cite{EMC91a,EMC87a,EMC84a}.
With the high energies reached in $ep$ collisions at HERA it is possible to
extend the studies of $x_F$ and $p_t^*$ distributions to larger values
of $W$, where the influence of perturbative QCD effects is expected to
be much larger and the final state hadron distributions should reflect
the dynamics of the subprocesses on the parton level.

In a recent analysis the scaled momentum and charged multiplicity
distributions of the hadronic final state were
measured in the current region of the Breit frame as a function of
the negative square of the four-momentum transfer, $Q^2$, and the
Bjorken-scaling variable $x$~\cite{breit}.
The charged particle spectra were observed to evolve
with $Q^2$ in a way similar as in \epem annihilation.
In this paper we study inclusive charged hadron production as a function of
$x_F$ and \ptm in the current region of the \gps\ frame.
The objective of the analysis is to investigate the influence of
perturbative QCD effects on the hadronic final state by
studying the $W$ dependence of these distributions in HERA $ep$
collisions and in fixed target DIS data.
The data are also compared with
\epem \ results as well as with predictions of Monte Carlo programs.
The comparison is also performed for a subclass of DIS events,
which are characterised by a rapidity gap between the observed
hadronic final state and the proton beam direction \cite{zeusgap},
and which are therefore candidates for diffractive scattering.

\section{ The experiment }
\subsection{ HERA }

The data were collected during the 1993 running period using the ZEUS detector
at the electron-proton collider HERA, where a 26.7 \gev \, electron
beam  and  a  820 \gev \, proton beam were brought to collision providing
an $ep$ centre-of-mass energy of 296 \gev.
84 bunches were filled for each beam and in addition 10 electron
and 6 proton bunches were left unpaired for background studies.
An integrated luminosity of 0.55~pb$^{-1}$ was collected.

 \subsection{ The ZEUS detector }

ZEUS is a multi-purpose magnetic detector which
has been described elsewhere \cite{green,sigp}. Here a brief description
is given which concentrates on those parts of the detector
relevant for the present analysis.

Charged particles are
tracked by the inner tracking detectors which operate in a
magnetic field of 1.43 T provided by a thin superconducting solenoid.
Immediately surrounding the beam pipe is the vertex detector (VXD) which
consists of 120 radial cells, each with 12 sense wires~\cite{vxd}.
The achieved resolution is $50~\mu$m in the central region of a cell
and $150~\mu$m near the edges.
Surrounding the VXD is
the central tracking detector (CTD) which consists of 72 cylindrical
drift chamber layers,  organised into 9 `superlayers'~\cite{ctd}.
These superlayers alternate between
those with wires parallel (axial) to the collision axis
and those inclined at a small angle to give a stereo view.
The hit efficiency of the CTD is greater than 95\% and the resolution
in transverse momentum  for full length tracks is $\sigma_{p_{T}}/p_{T}=
0.005\, p_T \bigotimes 0.016$ ($p_T$ in $\gev$), where $\bigotimes$ means
addition in quadrature.

The solenoid is surrounded by a high resolution  uranium-scintillator
calorimeter (CAL), which is divided into three parts:
forward\footnote{The proton beam direction is the forward $+Z$ direction.}
(FCAL), barrel (BCAL)
and rear (RCAL)  \cite{calor}. It covers 99.7\% of the solid angle.
Holes of $20\times 20$ cm$^2$ in the centre of FCAL and RCAL
accommodate the HERA beam pipe.
Each of the calorimeter parts is subdivided into towers which in turn are
segmented longitudinally into electromagnetic (EMC) and hadronic (HAC)
sections. These sections are further subdivided into cells, which are read
out by two phototubes each.

For measuring the luminosity as well as for tagging very small $Q^2$
processes, two lead-scintillator calorimeters are used~\cite{sigp,lumi}.
Bremsstrahlung photons emerging from the electron-proton interaction
point (IP) at angles $\theta_\gamma \le 0.5$ mrad with respect to the
electron beam axis hit the photon calorimeter at 107 m from the IP.
Electrons emitted from the IP at scattering angles
less than 6~mrad and with energies between 20\% and 90\% of the
nominal beam energy are deflected by beam magnets and
hit the electron calorimeter placed 35~m from the IP.

Two small lead-scintillator sandwich counters partially surround
the beam-pipe at the rear of the RCAL.
These counters were used to reject background produced by beam-gas
interactions with the incoming proton beam and to measure the timing
and longitudinal spread of both the proton and the electron beams of HERA.
Two layers of scintillation counters mounted on either side of an iron
veto wall, situated upstream of the detector, were also used to reject
background particles.

\section { Data taking conditions }

The ZEUS trigger is organised in three levels \cite{green} and reduces
the input event rate from the bunch crossing rate of 10~MHz to 3-5~Hz.
For DIS events, the first level trigger (FLT) requires at least
one of three conditions for energy
sums in the EMC calorimeter cells: the BCAL EMC energy exceeds 3.4 GeV; or
the RCAL EMC energy (excluding the innermost towers surrounding the beam pipe)
exceeds 2.0 GeV; or the RCAL EMC energy (including those towers) exceeds
3.75 GeV.

The second level trigger (SLT) rejects proton beam-gas events by
using the event times measured in the rear calorimeter cells.
The DIS trigger rate
of the SLT is about one-tenth the FLT DIS trigger rate.
The loss of DIS events at the SLT is negligible.

The third level trigger (TLT) has the full event information available and
applies physics-based filters. It requires tighter timing cuts to suppress
beam-gas background further and also rejects beam halo muons and cosmic muons.
The TLT selects DIS event candidates by calculating:
$$\delta~ =~ \sum_i E_i\cdot (1-\cos\theta_i)~~~>
         ~~~ 20~{\gev} ~~ - ~~2~E_{\gamma},$$
where $E_i$ and $\theta_i$ are the energy and the polar
  angle\footnote{ The proton beam direction is defined as the $Z$-axis in
   the HERA laboratory frame.}
 of the energy deposits in the calorimeter.
The summation runs over all calorimeter cells.
$E_{\gamma}$ is the energy measured in the photon calorimeter of the
luminosity monitor.
For fully contained DIS events $\delta \approx 2 E_e = 53.4$ GeV,
where $E_e$ is the energy of the incident electron.
Photoproduction events have low values
of $\delta$ compared to DIS events because the scattered electron
escapes in the hole of the calorimeter which contains the beam pipe.

For events with the scattered electron detected in the calorimeter, the
trigger is essentially independent of the DIS hadronic final state. The
trigger acceptance is greater than 97\% for $Q^2 > 10~\mbox{\gev}^2 $
and independent of $Q^2$~\cite{flt}.
A total of about $ 7 \cdot 10^6$ \ events passed the TLT and
was written to tape during the 1993 running period.

  \section{Event kinematics}

  In deep inelastic $ep$ scattering events the incoming electron couples to a
  $\gamma$ or a $Z$ (neutral current NC) or to a $W^{+}$ (charged current
  CC), which scatters off the proton. In the $Q^2$ range explored
  here, the contribution from $W$ and $Z$ exchange is negligible.
  The kinematic variables used to describe the inclusive DIS  process
  are defined in Table~\ref{kintab}.
\begin{table}[htb]
\centering
\begin{tabular}{|l|l|}
\hline
 \hspace*{1.5cm} {\bf Variable} & \hspace*{2.5cm} {\bf Description } \\
\hline
 & \\
 $l$ ($l'$) & Four-momentum of the incident (scattered) lepton \\
 $P$,$M_p$  &  Four-momentum of the proton and its mass \\
 $Q^2 = -q^2 = -(l-l')^2$  & Negative invariant mass squared of the exchanged
                             virtual \\
                           &  boson \\
 $\nu= (P \cdot q)/M_p$    & Energy of the
                             exchanged boson in the proton rest frame \\
 $ x=Q^2/(2P \cdot q)$     & Bjorken scaling variable \\
\hspace*{0.3cm} $ =Q^2/(2M_p\nu)$ & \\
 $ y = (P \cdot q)/(P \cdot l) $  &  Inelasticity parameter \\
$ W^2 = (P + q)^2 $  & Invariant mass squared of the hadronic final state \\
 \hspace*{0.7cm}$ = Q^2 \,(1-x)/x + M_p^2 $
& \\
\hline
\end{tabular}
\caption{ Definition of the variables used to describe the
          kinematics of the inclusive DIS process }
\label{kintab}
\end{table}

  The ZEUS detector is almost hermetic, allowing the kinematic variables
  $Q^2,x$ and $y$ to be reconstructed in a variety of ways using combinations
  of electron and hadronic system energies and angles.
  In the analysis presented here the double angle method (DA) was chosen,
  in which the scattered electron angle and the angle $\gamma_H$ is used
  \cite{f2:17}. In the naive quark parton model $\gamma_H$ corresponds to
  the angle of the scattered massless quark in the laboratory frame.
  The variable $y$ is determined according to the Jacquet-Blondel method
  \cite{jb} and is denoted by $y_{JB}$.

  The four-momentum of the scattered electron needed to calculate
  the Lorentz  boost to the \gps\ frame, is reconstructed from its
  polar and azimuthal angle, $\theta_e,\phi_e$.
  The scattered electron energy $E^{\prime}_{DA}$, used in the boost,
  is computed by the double angle method:
\bes
 \label{equedbl}
    E^{\prime}_{DA} = Q^2_{DA}/(2E_e\,(1+cos{{\theta}_e})) \ ,
\ees
   where $E_e$ is the energy of the incident electron and $Q^2_{DA}$ is
   given by:
\bes
 \label{equeqqdbl}
    Q^2_{DA} = \ 4~E^2_e \cdot
               { {\sin{\gamma_H} \ (1+\cos{\theta_e)}} \over
                 {\sin{\gamma_H} + \sin{\theta_e} - \sin{(\gamma_H+\theta_e)}}
               } \ .
\ees

  The variables $x_F$ and $p_t^*$ describe the kinematics of the hadrons
  in the \gps:
\be
\label{equxf}
  x_F = p_{l}^* / |p^*_{l,max}| \,\,\, = \,\, 2 p_{l}^* /W \,\, ,
\ee
  where $p_{l}^* $ is the projection of the hadron momentum vector
  onto the direction  of the virtual photon and $|p^*_{l,max}|$ is the maximum
  value of $p_{l}^* $.
  The  hadron momentum component   perpendicular to the virtual photon axis
  is denoted by $p_t^{*}$.

 \section{ Data selection }

\subsection{ Event selection }

The offline selection of DIS events was similar to that described in
 earlier publications \linebreak[4]
 (e.g. \cite{breit,zeusf293,lrgefl}).
Scattered electron candidates
were selected by using the pattern of energy deposition in the
calorimeter. The electron identification algorithm was
tuned for purity rather than for efficiency. The purity is defined
as the number of electrons generated and reconstructed in a bin
divided by the total number of electron candidates measured in the bin.
In studies with Monte Carlo DIS events and test beam data the
purity was estimated to be $\geq 96~\%$ for $E^{\prime}_{DA} \ge 10~{\gev}$.

The requirements for the final event selection were:
\begin{itemize}
\item
   $E^{\prime}_{DA} \ge 10~{\gev}$,
   to minimise beam gas background contamination;
\item
   $Q^2_{DA}\geq 10$ GeV$^2$;
\item
   $y_e\leq 0.85$,
   to reduce the photoproduction background, where $y_e$ is the scaling
   variable $y$ as determined from the energy and polar angle of the
   scattered electron;
\item
   $y_{JB}\geq 0.04$, to guarantee sufficient accuracy for the $DA$
   reconstruction method;
\item $\delta = \sum_i E_i(1 - \cos\theta_i)  \ge 35$~GeV,
   where the sum runs over all calorimeter cells.
   For fully contained events $\delta \approx 2 E_e = 53.4$~GeV.
   This cut is used to remove photoproduction events and to control
   radiative corrections.
\end{itemize}
Furthermore we required:
\begin{itemize}
\item
   a primary vertex position, determined from VXD and CTD tracks, in the
   range \linebreak $-50 \leq Z_{vtx} \leq  40~\cm$;
%
\item
  the impact point $(X,Y)$ of the scattered electron in the RCAL to lie
  outside a square of $32 \times 32$ cm$^2$ centered on the beam axis, to
  ensure that the electron is fully contained within the detector and
  its position can be reconstructed with sufficient accuracy.
\end{itemize}
After these cuts, the remaining photoproduction background was
estimated  to be $\simeq$ 1\%.
The contamination from beam-gas background was estimated to be below 0.5\%
as calculated from unpaired electron and proton bunches.
Finally, QED Compton scattering events and residual cosmic and
beam-related muons were rejected by algorithms, which identify
this types of events by their pattern of energy deposits in the
calorimeter cells.

A total of 26100 events was selected by the above cuts.
Of these events about 10\% \cite{zeusgap}
contain a large rapidity gap in the hadronic final state.
They are characterised by
$\eta_{max}<1.5$, where $\eta_{max}$ is the pseudorapidity of
 the most forward calorimeter cluster
in the event, relative to the proton direction. The
pseudorapidity is defined by $\eta = -\ln{(\tan{(\theta/2)})}$ and
a cluster is an isolated set
of adjacent calorimeter cells with summed energy above 400 MeV.
  This sample is called the `large-rapidity-gap' (LRG) event sample.
  The remaining events are denoted by `non-rapidity-gap' (NRG) events.
  The invariant mass of the hadronic final state
  excluding the scattered proton in the LRG events is
  calculated from the energy deposits measured in the calorimeter
  (excluding the electron cluster) by
  $M_X = \sqrt {\sum_{had} (E^2 - p^2_X - p^2_Y - p^2_Z )} $. The values of
  $p_X,~p_Y$ and $p_Z$ are the cell energies $E$ projected on the axes
  of the HERA laboratory frame. The polar angles of these pseudovectors
  are calculated from the geometric centres of the cells and the
  primary event vertex position. The measured value of $M_X$ is corrected
  to the hadron level as described in section~\ref{acceptance}.

\subsection{ Track reconstruction and selection }

Tracks were recognised and fitted using two programs which were
developed independently and follow different strategies for pattern
recognition and track fitting.
For the results shown in this paper the first approach is used and
the second method was used  for estimating the  systematic error.

In the first approach the track finding algorithm starts with
hits in the  outermost axial superlayers of the CTD.  As the
trajectory is followed inwards to the beam axis,
more hits from the axial wires of the CDT and of the VXD are incorporated.
A circle is fitted in the $XY$ projection and is used for the pattern
recognition in the stereo superlayer pattern.
The momentum vector is determined in a 5-parameter helix fit.

The second track finding program is based on the Kalman filtering
technique \cite{kalman}. Seed tracks found  in the outer
layers of the CTD are extended  inwards and points are added as wire
layers of the CTD are crossed. The track
parameters at each step are updated using the Kalman method.
In the second step a Kalman fit to the points found in the pattern
recognition phase is performed taking into account non-linear corrections
to the measured drift time.
Following the reconstructed CTD track inwards, CTD and VXD hits
are associated with the track.
The VXD track segments are merged with the CTD tracks
using the Kalman filtering algorithm.

Multiple Coulomb scattering in the beam pipe
and in the walls of the VXD and CTD
were taken into account in the evaluation of the covariance matrix.
The vertex fit is performed with the fitted tracks using the
perigee parameterisation \cite{billoir}. The vertex position is evaluated and
the track parameters are calculated at
  the vertex.

  Only tracks which are  associated with the primary vertex have been selected
  for this analysis.
  The tracks are required to have $p_{t,lab}> 0.2 \gev$
  and a polar angle in the HERA laboratory frame in the range of
  $ 25 \deg < \theta < 155 \deg $.
  This is a region of the CTD,
  where the detector response and systematics are best understood.
  For tracks defined by these cuts the track
  reconstruction efficiency is $\simeq$~95\%.

  The scattered electron was removed from the track sample by rejecting
  those tracks which match the cluster in the calorimeter assigned to the
  scattered electron by the electron finding algorithm.
  Only tracks which reach at least the third superlayer and hence have a
  projected length in the plane perpendicular to the beam axis of more than
  30~cm are kept to achieve the required transverse momentum resolution.
  For $\theta > 150\deg$ the efficiency for identifying the scattered
  electron by matching the CTD tracks to energy deposits in the calorimeter
  decreases rapidly due to the limited acceptance and resolution of the
  CTD in the very rear part of the detector. Therefore the upper cut on
  $\theta$ of the hadrons considered in the analysis was further tightened
  to $150\deg$.

  Due to the cuts in $\theta$ and $p_{t,lab}$ the analysis
  in the \gps\ is restricted to the range
  $10 < Q^2 < 160 \gevv$ and $75 < W < 175 \gev $, where the
  acceptance for charged hadrons is larger than $ 60\%$.

\begin{figure}[tp]
 \epsfxsize=18cm
 \epsfysize=18cm
 \epsffile{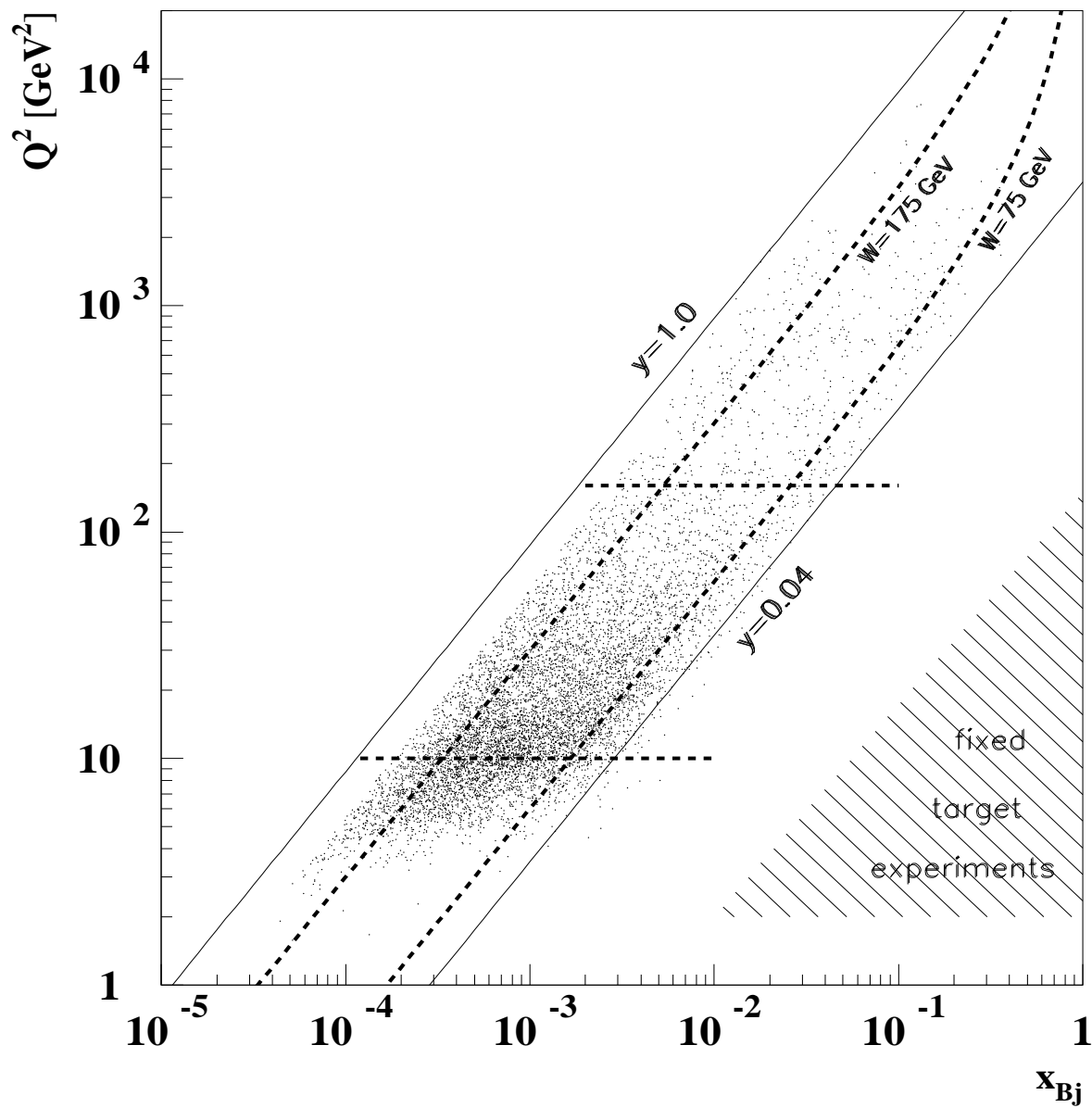}
 \caption{
    Population of the $Q^2$-$x$ plane by the DIS events selected
    for this analysis. For the sake of clarity only 1/3 of the DIS event
    sample is shown in the scatter plot. Charged hadron distributions are
    investigated for $10 < Q^2 < 160 \gevv$ and $75 < W < 175 \gev $
    (dashed lines). The approximate kinematic region covered by the
    fixed target experiments is also indicated.
          }
 \label{qx}
 \end{figure}

  In Fig.~\ref{qx} the distribution of the selected events in the
  $Q^2$-$x$ plane is shown. For comparison the kinematic region which has been
  investigated in fixed target experiments is also shown.

  \section{ Acceptance correction }
  \label{acceptance}

  \subsection{ Monte Carlo simulation }

  The measured distributions are corrected for detector effects such as
  acceptance  and resolution.   For that purpose the
  hadronic final state from DIS was modelled using two different sets
  of Monte Carlo generators, the first for the description of the
  non-rapidity-gap events and the second to model the
  large-rapidity-gap events.

  Events from NRG DIS processes were generated using two alternative
  Monte Carlo models: a) the combination of the LEPTO 6.1
  \cite{lep60} and the ARIADNE 4.0 Monte Carlo program \cite{cdm,and83}
  (CDMBGF) and b) LEPTO 6.1  with the option of combined matrix element
  and parton   shower calculation   (MEPS).
  The fragmentation was simulated using the
  LUND string model \cite{string80} as implemented in JETSET \cite{jetset}
  (see Table~\ref{mctab}).

  Both models were interfaced to the  program
  HERACLES \cite{heracl}, which computes the electro-weak radiative corrections
  for DIS events.
  In the case of hard QED Bremsstrahlung the four-momentum vector
  of the   virtual photon which probes the proton  
  is significantly
  different from the virtual photon momentum reconstructed from the momenta
  of the incident  and scattered lepton.
  In this case the $x_F$ and $p_t^*$ distributions are also distorted and
  have to be corrected for this effect.
  In this analysis, however, the virtual photon momentum was reconstructed
  using the double angle method, which is insensitive to radiative effects.
  Events with hard QED initial state Bremsstrahlung photons
  ($E_{brems}~\sgeq~7 \gev$) are rejected by the cut on $\delta>35 \gev$
  (see section~5.1).
  Monte Carlo calculations show that the QED radiative
  corrections  
  are $ 5-10 \%$.

  For both Monte Carlo simulations the $MRSD_{-}'$ parameterisation
  of the parton densities in the proton
  was chosen \cite{mrsd93}, which gives a reasonable description of the
  structure function measured at HERA \cite{f2zeus,f2h1}.

  The properties of LRG events are characteristic of diffractive
  interactions \cite{zeusgap}.
  Two Monte Carlo  event samples have been used to model the hadronic final
  state of LRG events. The first was generated using
  the   POMPYT  Monte Carlo program \cite{pomp}, which is based on
  a factorisable   model for high energy diffractive processes. Within the
  PYTHIA   \cite{pyt} framework, the incident proton emits a pomeron, whose
  constituents take part in a hard scattering process with the virtual photon
  or its constituents. The structure of the pomeron is assumed to be
  described by either a hard or a soft quark density function $f(\beta)$,
  where $\beta$ denotes the fraction of the pomeron momentum carried by the
  quark.

  The second sample was generated following the model of Nikolaev and
  Zakharov (NZ) \cite{nikzak}, which was interfaced to the Lund fragmentation
  scheme \cite{ada}. In the NZ model it is assumed that the exchanged virtual
  photon fluctuates into a $q\bar{q}$ pair, which interacts with a colourless
  two-gluon system emitted by the incident proton.
   Both diffractive Monte Carlo samples were
  generated with default parameter settings. QED radiative processes were not
  simulated for these events. With the event selection cuts described in
  section~5, however, the QED radiative corrections are expected
  to be  of the same size as for the NRG events.
\begin{table}  
\centering
\begin{tabular}{|l|l|}
\hline
 Acronym & Description \\
 \hline
 & \\
 QPM        &  Quark parton model + string fragmentation only \\
 CDM        &  Colour dipole model \protect \cite{cdm,and83} \\
 MEPS (*)   &  Parton shower \protect \cite{lep60} matched to
               complete ${\cal{O}}(\alpha_s)$ matrix element \\
            &  calculation (ME) \\
 CDMBGF (*) &  Colour dipole model combined with
               complete ${\cal{O}}(\alpha_s)$ matrix element \\
            &  calculation for the BGF process (ME) \\
 POMPYT (*) &  Model for diffractive DIS
               (assuming factorisation of the pomeron \\
            &  flux and the pomeron structure function)
               \protect  \cite{pomp} with  \\
            &  a hard quark density function for the pomeron
               $ \propto [\beta(1-\beta) \label{equfpomh} ]$ \\
            &  or a soft quark  density function for the pomeron
               $ \propto [(1-\beta)^5 \label{equfpoms}]$  \\
 NZ (*)     &  Model for diffractive DIS (non factorisable ansatz)
               \cite{nikzak} \\
\hline
\end{tabular}
\caption{ Acronyms for the DIS models used in this report.
   For those generators marked by an asterisk, event samples have also been
  processed by the detector simulation and data reconstruction program.
  In all models  the LUND string
  fragmentation model is used \protect \cite{string80,jetset}.
}
\label{mctab}
\end{table}

  Event samples produced by the Monte Carlo generators marked in
  Table~\ref{mctab} by an asterisk were also  processed
  by the ZEUS detector simulation program, which is based on GEANT 3.13
  \cite{brun} and which incorporates the detector and trigger simulation.
  Events fulfilling the trigger conditions were then passed through
  the standard ZEUS offline reconstruction program.

  The predicted  $\eta_{max}$ distribution for non-diffractive
  DIS events falls exponentially for $\eta_{max}< 4$,
  whereas for diffractive events this distribution is approximately flat.
  Calculations with the CDMBGF Monte Carlo model show that
  the fraction of non-diffractive DIS events  with $\eta_{max}< 1.5$
  is about 5\% \cite{zeusgap}.
  The distributions  for the LRG event  sample defined by $\eta_{max}< 1.5$
  have been corrected with POMPYT
  and those for the NRG events have been corrected using the
  CDMBGF Monte Carlo program interfaced  to HERACLES.
  Note that the results are not corrected for the selection inefficiency of
  the $\eta_{max}$ cut.

  \subsection{ Data correction procedure }

  The measured hadron multiplicity distributions are distorted with respect
  to those of the true hadronic final state due to trigger biases,
  event and track  selection cuts and the acceptance and resolution
  of the detector. The output of the trigger and detector simulation program
  together with the samples produced by the different event generators
  have been used to estimate the distortion of the distributions and to
  correct for them by multiplying the measured distributions by a correction
  function $c(v)$ in each bin of $Q^2$ and $W$, where $v$ is the
  hadron variable under study and $c(v)$ is calculated as a bin-by-bin
  ratio:
 \bes
 \label{equacc}
 c(v) = \left( \frac{1}{N_{evt}} \frac{\Delta N_{had}(v)}{\Delta v}
\right)_{gen} \left. \right/
 \left( \frac{1}{N_{evt}} \frac{\Delta N_{had}(v)}{\Delta v}    \right)_{rec}
   \,\, .
  \ees
  The subscripts $gen$ and $rec$ refer to the quantities as given
  by the event generator programs and the reconstructed quantities
  from the output of the detector simulation program, respectively.
  The number of events in a bin of $Q^2$ and $W$
  is denoted by $N_{evt}$; $\Delta N_{had}$ is the number of
  hadrons in a bin of $v$.
  The generated hadron distributions do not include the charged particle decay
  products of $K^0$'s and $\Lambda$'s and of weakly decaying particles with
  a lifetime  $> 10^{-8}$s.
  For the expression in the numerator
  events  and hadrons are sorted in bins of the generated kinematic
  variables and for the denominator in bins of the reconstructed variables.
  In this way the distributions have been corrected for
  losses of events and hadrons as well as for the
  effects of  event migration, finite resolution and trigger biases.


  The bin size in the hadron variables $v$
   was chosen to be comparable with the estimated resolution in $v$ and it was
  checked that the correction factor neither deviates by more than 40\% from
  unity nor depends strongly on $v$ \cite{heiko}.
   For models which adequately describe
  the data, the dependence of the correction factors on the
  model input was found to be small.
  The difference in $c(v)$ for different models was included in the
  systematic error.

  The mean square of $p_t^{*}$ ($<p_t^{*\,2}>$) was corrected by:
\bes
      <p_t^{*\,2}>        = <p_t^{*\,2}>_{meas} \,
   \frac{<p_t^{*\,2}>_{MC,gen}} {<p_t^{*\,2}>_{MC,rec}} \,\, ,
\ees
 where $<p_t^{*\,2}>_{meas}$ is the mean value of $p_t^{*\,2}$ determined
  from the uncorrected data. The terms in the correction factor
  are defined as in equation \ref{equacc}.
  This method of correction is numerically more stable than the determination
  of $<p_t^{*\,2}>$ from the acceptance corrected $p_t^{*\,2}$ distributions.

  The following sources of systematic uncertainties were studied:
\begin{itemize}
\item
   The model dependence of the correction factors $c(v)$ was estimated
   using two different models for the NRG and LRG event samples each.
   The CDMBGF and MEPS models were used to correct the NRG event sample
   and for the LRG event sample the POMPYT model with a hard quark
   density function (see Table~\ref{mctab}) and the NZ model were used.
   The relative systematic error of $1/N_{evt} \cdot dN_{had}/dx_F$ is
   $\sim 3\% $ and the one of
   $1/N_{evt} \cdot dN_{had}/d<\!\!p_t^{*\,2}\!\!>$ is $\sim 7\% $.
\item
  The analysis was done using two different strategies for track finding
  and vertex fitting as described in section 5.2.
  The difference of the corrected $x_F$ and $p_t^{*}$ distributions
  obtained with both programs is used as an estimate
  of the systematic error from the track reconstruction.
  The relative systematic error of $1/N_{evt} \cdot dN_{had}/dx_F$
  is $\sim 10\% $ and the one of
  $1/N_{evt} \cdot dN_{had}/d<\!\!p_t^{*\,2}\!\!>$ is $\sim 4\% $.
\item
  Systematic uncertainties in the determination of the four-momentum
  of the virtual photon may induce a systematic error in the
  hadron distributions measured as a function of
  $x_F$ and $p_t^{*}$. The size of this systematic error was estimated
  from Monte Carlo events by using the generated four-momentum
  of the virtual photon rather than the reconstructed four-momentum.
  The Lorentz transformation with the generated values
  was then used to calculate the momenta
  of the reconstructed final state particles in the \gps\
  and  these values were compared to those obtained via the reconstructed
  virtual photon momentum.
  The relative systematic error of $1/N_{evt} \cdot dN_{had}/dx_F$
  is $\sim 7\%$ and the one of
  $1/N_{evt} \cdot dN_{had}/d<\!\!p_t^{*\,2}\!\!>$ is $\sim 5\% $.
\item
  The sensitivity of the measurements on the track selection criteria
  has been investigated.
    The cut in the polar angle of the tracks   was varied between
  $20\deg$ and  $33\deg$
  and/or it was required   that superlayer 5 instead of
  superlayer 3 has to be reached by   the   track. The requirement of
  a minimum hadron momentum  transverse to the beam direction in
  the laboratory frame, $p_{t,lab}$, was omitted.
  No significant changes in the results
   ($<1\%$)  have been observed.
\item
  The effect of a possible misestimation of the momentum resolution in the
  detector simulation program was studied by evaluating the correction function
  with a resolution of the measured transverse momentum
  artificially increased by
  100\%. The size of this effect on \ptm and $x_F$ was  smaller than  1\%.
\end{itemize}

  The contributions of the above effects to the systematic error have been
  added in quadrature and are shown together with the statistical errors
  of the results in the tables and figures.

\begin{figure}[tp]  
 \epsfxsize=18cm
 \epsfysize=18cm
 \epsffile{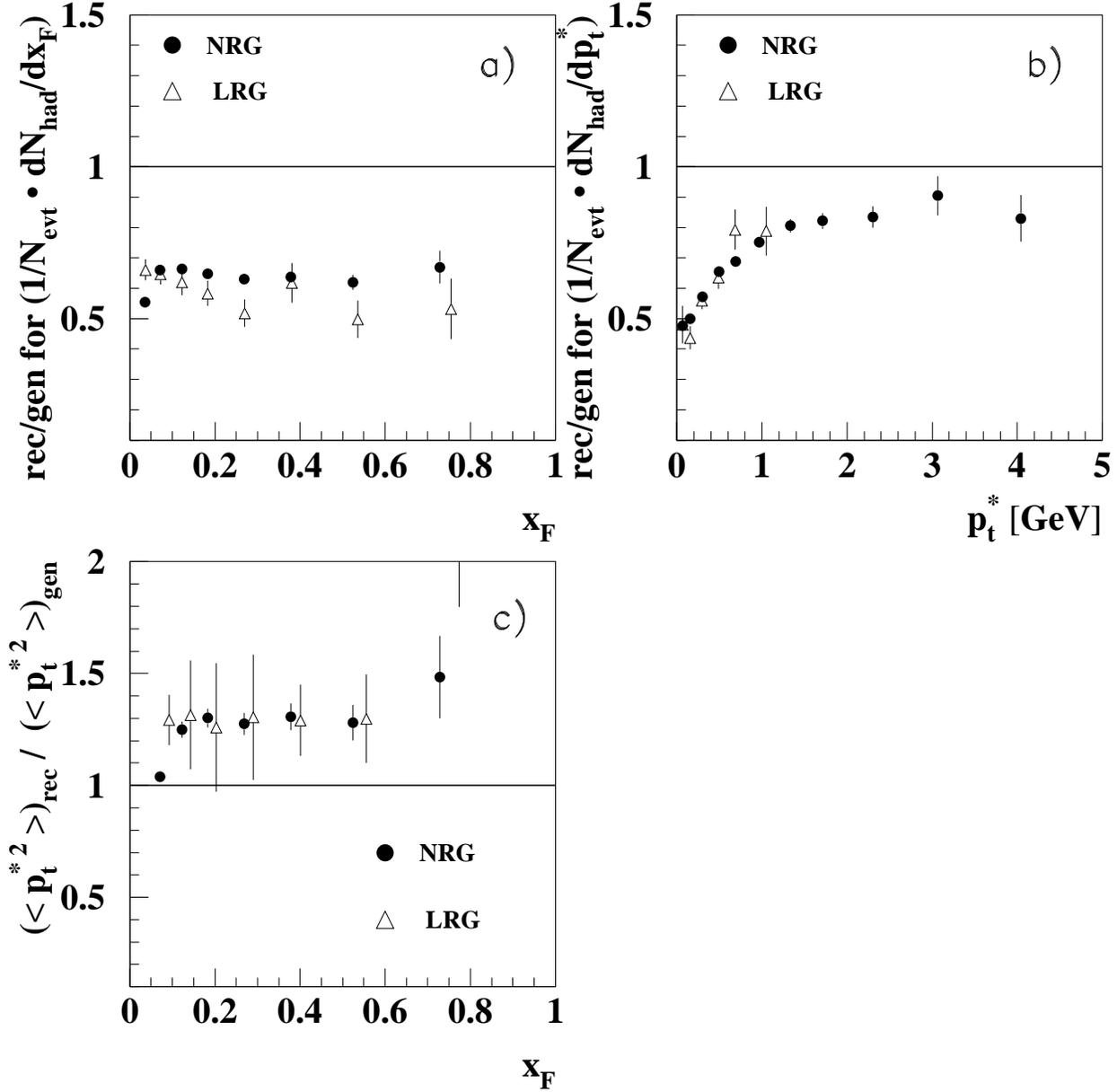}
 \caption{
     Inverse of the correction functions $c(v)$ for a) the $x_F$
     and b) the $p_t^{*}$ distribution
     in the range of $10<Q^2<160 \gevv$ and $75<W<175 \gev$,
     which are used to correct the NRG event sample (full points)
     and the LRG event sample (triangles).
     The inverse of the correction function for $<p_{t}^{*\,2}>$ as a
     function of $x_F$ in the same range of $Q^2$ and $W$ is shown in
     c).
         }
 \label{accxfpt}
 \end{figure}

  The shape  of the correction factors to be applied to
  the measured hadron distributions of $x_F$ and $p_t^{*}$ as well as
  to $<p_t^{*\,2}>$ is shown in
  Fig.~\ref{accxfpt} separately for NRG and LRG events.
  The size of the correction for both event classes
  is very similar.

 \section{ Results }

  \subsection{ $x_F$ and $p_t^{*}$ distributions in NRG events }

\begin{figure}[tp]
 \epsfxsize=18cm
 \epsfysize=18cm
 \epsffile{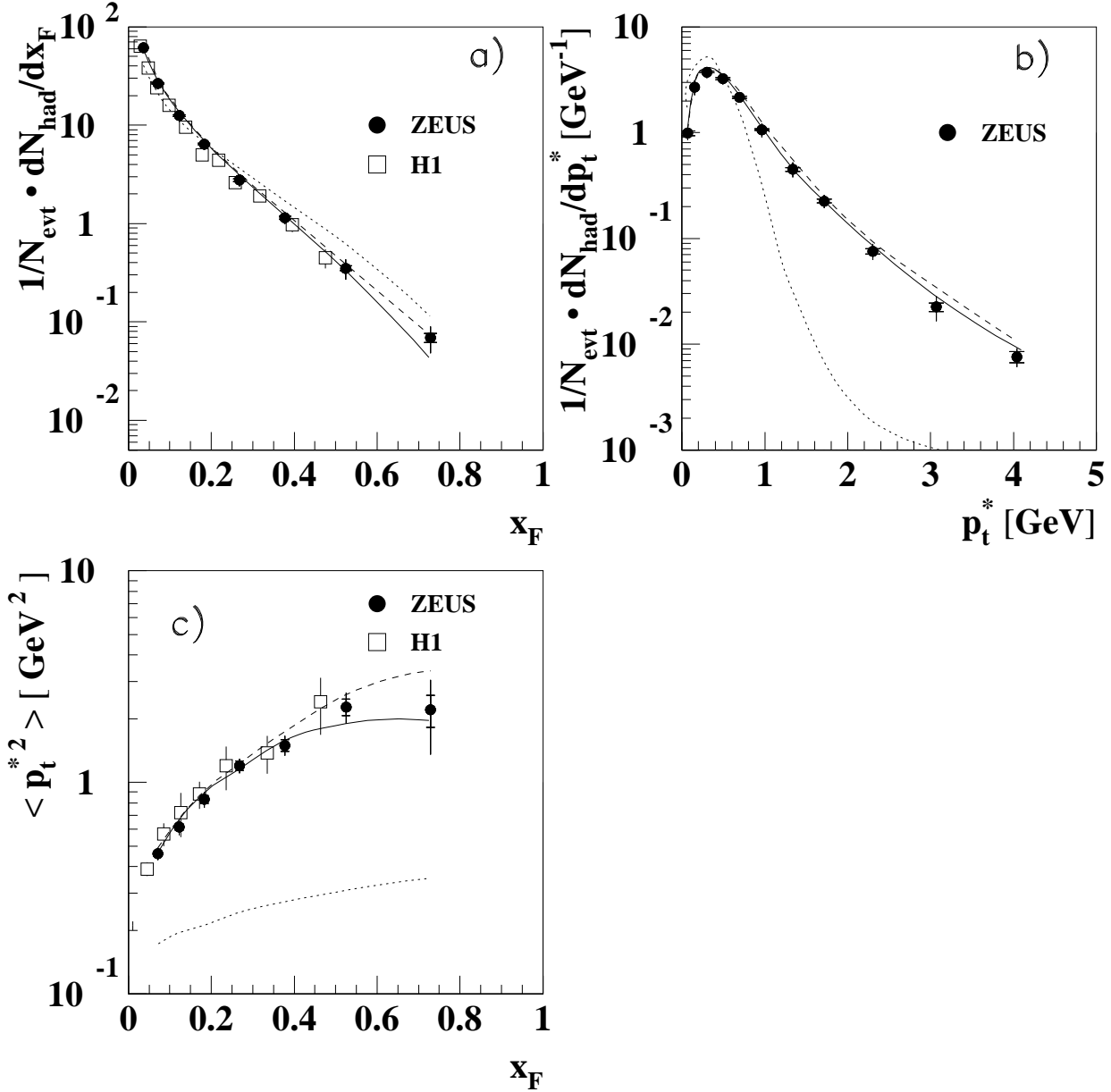}
 \caption{
     Differential charged hadron multiplicities for NRG DIS events
     normalised by the number of events as a function of
     a) $x_F$ and b) $p_t^{*}$ for $x_F > 0.05$.
     c) $<\!p_t^{*\,2}\!>$  as a function of $x_F$.
     For all plots the events are  in
     a range $10 \le Q^2 \le 160 \gevv$ and $75 \le W \le 175 \gev$.
     The predictions of two DIS Monte Carlo models including QCD
     processes are shown: the MEPS model (solid curve) and the CDMBGF model
     (dashed curve). The prediction of the QPM
     is given by the dotted curve.
     The results of this analysis in a) and c)
     are also compared to measurements of
     the H1 collaboration \protect \cite{hfsh1}.
         }
 \label{xfptnrg}
 \end{figure}

 First the $x_F$ and $p_{t}^{*}$ distributions of charged hadrons
 in NRG events are  discussed. In Fig.~\ref{xfptnrg}a
 the $x_F$ distribution at $<W> = 120 \gev$ and $<Q^2>=28 \gevv$
 is compared with different models for hadron production in DIS.
 The $x_F$ distribution falls steeply with increasing $x_F$. The
 results from the H1 experiment \cite{hfsh1} agree well with this
 measurements.  The data agree with those models, in which
 higher order QCD processes are included, such as MEPS (solid line) and
 CDMBGF (dashed line), but not with the naive quark parton model
 (QPM) (dotted line).

  In Fig.~\ref{xfptnrg}b the $p_t^{*}$ spectrum, which is integrated over
  $x_F > 0.05$ for the study of the current jet fragmentation,
  is compared with the same model calculations.
  The QPM model predicts a much steeper $p_t^{*}$ distribution than the
  data show, whereas the MEPS model agrees well with the data.
  However, for closer investigation it is advantageous to take the
  mean square of $p_t^{*}$, $<p_{t}^{*\,2}>$, a quantity which is more
  sensitive to the behaviour of the tail of the $p_t^{*\,2}$ distribution.

  Figure~\ref{xfptnrg}c shows the $<\!p_t^{*\,2}\!>$ distribution
  as a function of $x_F$ for $x_F \ge 0.05$.
  In any model, which allows for a transverse momentum of the partons,
  the rise of $<p_t^{*\,2}>$ with increasing $x_F$ is expected
  because a hadron with a higher value of $x_F$ carries also
  a larger fraction of the transverse momentum of the primary parton.
  Again the MEPS and CDMBGF models describe the data while the QPM strongly
  underestimates the value of $<\!p_t^{*\,2}\!>$.

  In Fig.~\ref{xfptnrg}a,c the results from the H1 experiment are also
  shown \cite{hfsh1}. The differential hadron multiplicities measured
  by ZEUS are listed in Table~\ref{nrgtab}.

  \subsection{ $x_F$ and $p_t^{*}$ spectra  in  LRG events }

\begin{figure}[tp]
 \epsfxsize=18cm
 \epsfysize=18cm
 \epsffile{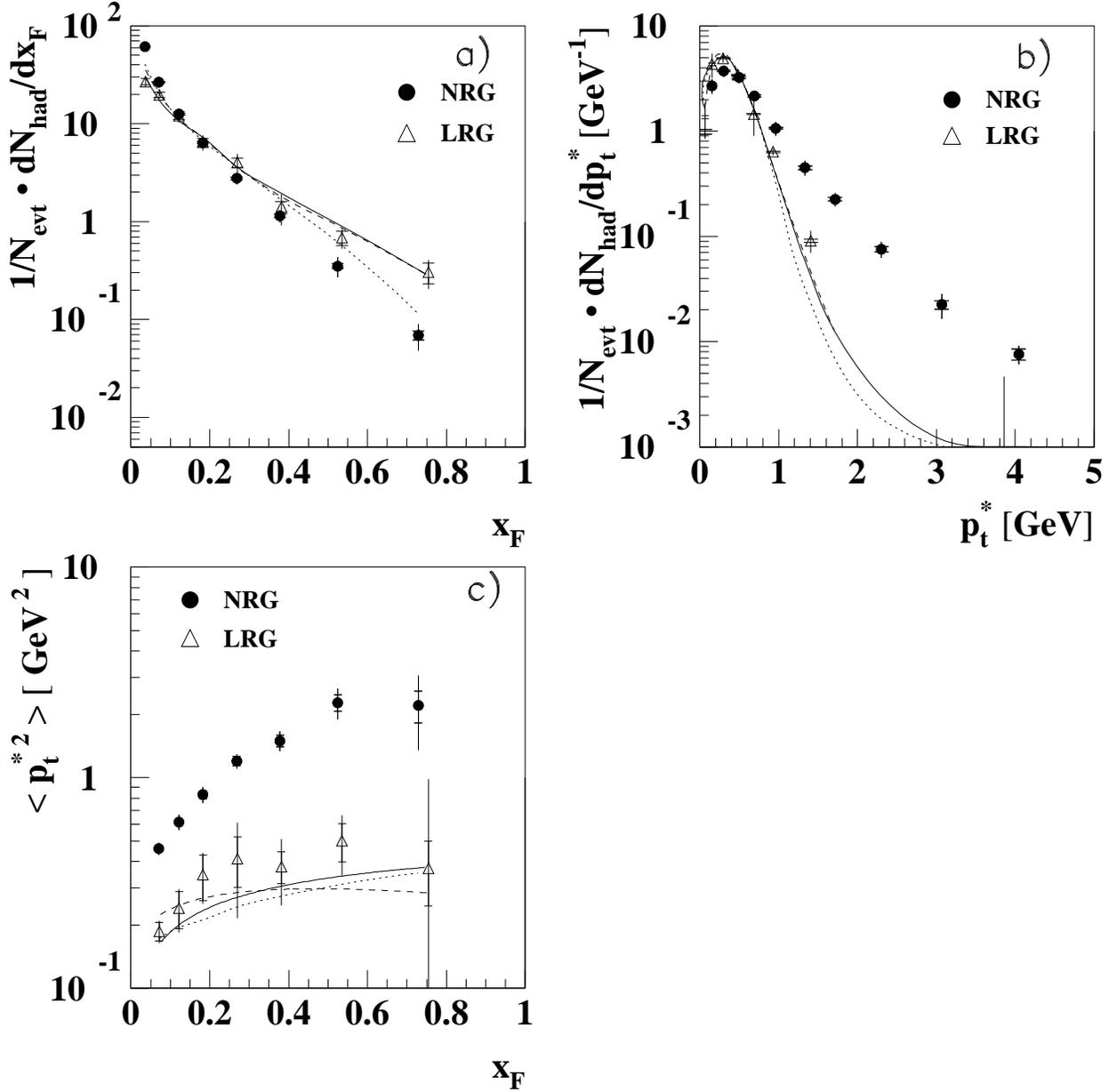}
 \caption{
    Charged hardon distributions for $10 \le Q^2 \le 160 \gevv$ and
    $ 75 \le W \le 175 \gev$ ($<Q^2>=28 \gevv$ and $<W>=120 \gev$).
    a) The $x_F$ distribution, b) the $p_t^{*}$ distribution for
    $x_F > 0.05$ and c) $<p_t^{*\,2}>$ as a function of $x_F$ are
    presented separately for NRG and LRG events. In all three figures
    the curves represent the results of the following model predictions:
    solid curve: POMPYT with a hard pomeron structure function
    (see Table~\protect\ref{mctab}); dashed curve: model of Nikolaev
    and Zakharov; dotted curve: QPM.
        }
 \label{xfptdiff}
 \end{figure}

  The $x_F$ and $p_t^{*}$ distributions from charged hadrons as well as
  $<\!p_t^{*\,2}\!>$ as a function of $x_F$ are shown in Fig.~\ref{xfptdiff}
  separately for the samples of LRG and NRG events. The values for the
  LRG events are tabulated in Table~\ref{lrgtab}. The value of
  $<W>$ is similar for both event samples, whereas $<Q^2>$
  for the LRG events   is lower by 30\% than for the NRG events.
  The $x_F$ distribution for the LRG events is falling less steeply when
  compared to that of the NRG events.
  The LRG data in Fig.~\ref{xfptdiff} are reasonably well described
  by the POMPYT (solid line) and the NZ (dashed line) models for
  diffractive DIS with the $\eta_{max}$ cut applied.
  The QPM prediction for the $x_F$ distribution of DIS events,
  shown by the dotted line in Fig.~\ref{xfptdiff}a, is slightly steeper
  than the $x_F$ distribution for LRG events.

  The $p_t^{*}$ spectrum of LRG events is significantly less
  broad than that for the rest of the DIS events (Fig.~\ref{xfptdiff}b).
  This effect is highlighted  in Fig.~\ref{xfptdiff}c.
  The mean values of $p_t^{*\,2}$ in events with a large rapidity gap
  are smaller than for the NRG events by a factor of 2--5.
  From a comparison with DIS model calculations with and without  simulating
  QCD radiation processes, it is found that the $<\!p_t^{*\,2}\!>$ values
  for LRG events resemble those for DIS events with only a small amount of
  gluon radiation.
  This observation is in good agreement with ZEUS results from the
  analysis of the energy   flow \cite{lrgefl}.
  However, $<p_t^{*\,2}>$ in LRG events is somewhat larger than
  predicted by the QPM (see dotted line in Fig.~\ref{xfptdiff}c),
  indicating that there is a non-zero contribution
  of higher order QCD processes in this class of events, too. This is
  confirmed by the observation of DIS events with a large rapidity gap which
  exhibit a two-jet structure \cite{lrgjet}.
  The   model calculations for diffractive $ep$ scattering slightly
  underestimate the measured values of  $<p_t^{*\,2}>$.



\begin{figure}[tp]
 \epsfxsize=18cm
 \epsfysize=10cm
 \epsffile{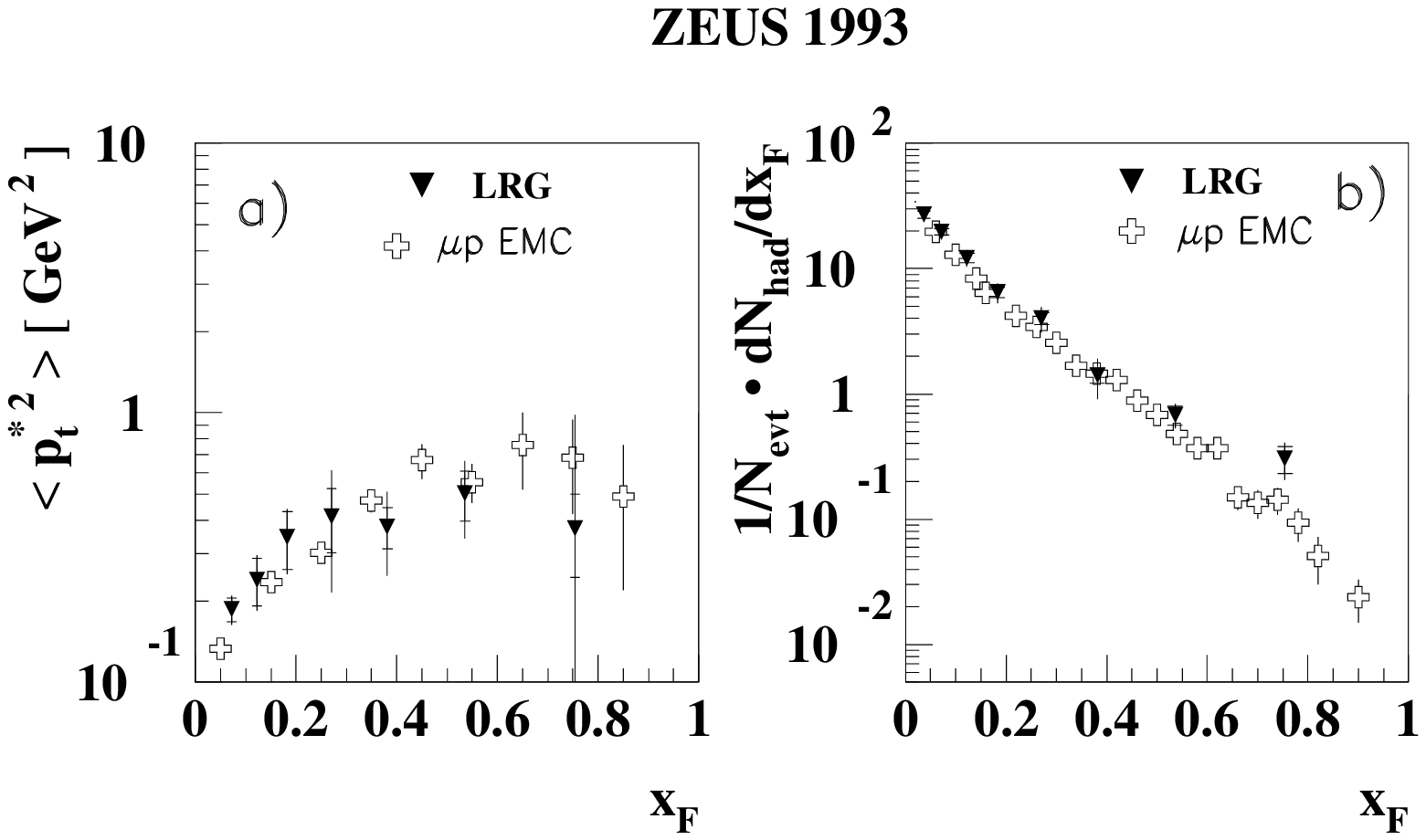}
 \caption{
    Comparison of a) $<\!p_t^{*\,2}\!>$ as a function of $x_F$
    and b) the $x_F$ distribution for the LRG event sample (ZEUS) and
    DIS at low energy (EMC, $<W> = 14 \gev$). The mean value of $M_X$ for
    the LRG event sample is $<M_X>=8 \gev$.
         }
 \label{pxfptlemcp}
 \end{figure}

  The inclusive distributions of LRG events have been found to have the
  properties of
  a diffractive interaction of a highly virtual photon with a proton
  \cite{zeusgap}. Diffractive interactions in hadron-hadron reactions and
  photoproduction have been successfully described in the framework of Regge
  theory by the exchange of a pomeron \cite{dla84}. Several models have
  been developed to describe this reaction in terms of parton
  interactions (e.g. \cite{ing84,nik92}).
  In this context it is interesting to test the hypothesis
  that the diffractive  DIS process can be viewed as the `emission' of
  a pomeron from the proton, which carries the fraction $x_{pom}$ of the
  proton momentum, and a subsequent deep
  inelastic $\gamma^*$ pomeron scattering, which occurs at a higher value
  of $x'=\frac{x}{x_{pom}}$.
  In this picture     
  the relevant scale for the invariant mass of the hadronic final
  state should be given by $M_X$ and not by $W$.
  In Fig.~\ref{pxfptlemcp}a  $<\!p_t^{*\,2}\!>$ as a function of $x_F$
  from the LRG events is compared with the results of a fixed target
  DIS experiment \cite{EMC91a}, where the invariant mass of the total
  hadronic final state ($<\!W\!>= 14 \gev$) is only slightly higher than
  the invariant mass of the hadronic final state observed in the LRG events
  ($<\!M_X\!> = 8 \gev$). The values of  $<\!p_t^{*\,2}\!>$ and the $x_F$
  distribution for both
  event samples agree reasonably well. This result supports the hypothesis
  that the transverse momentum space for the particle production is similar
  to DIS, where the scale of the invariant mass is given by $M_X$ rather
  than by $W$.

  \subsection{ $W$ and $Q^2$ dependence of $x_F$ and $p_t^*$ spectra }

\begin{figure}[tp]
\centerline{
  \hspace*{0.7cm}
  \epsfxsize=14.0cm
  \epsfysize=19.0cm
  \epsffile{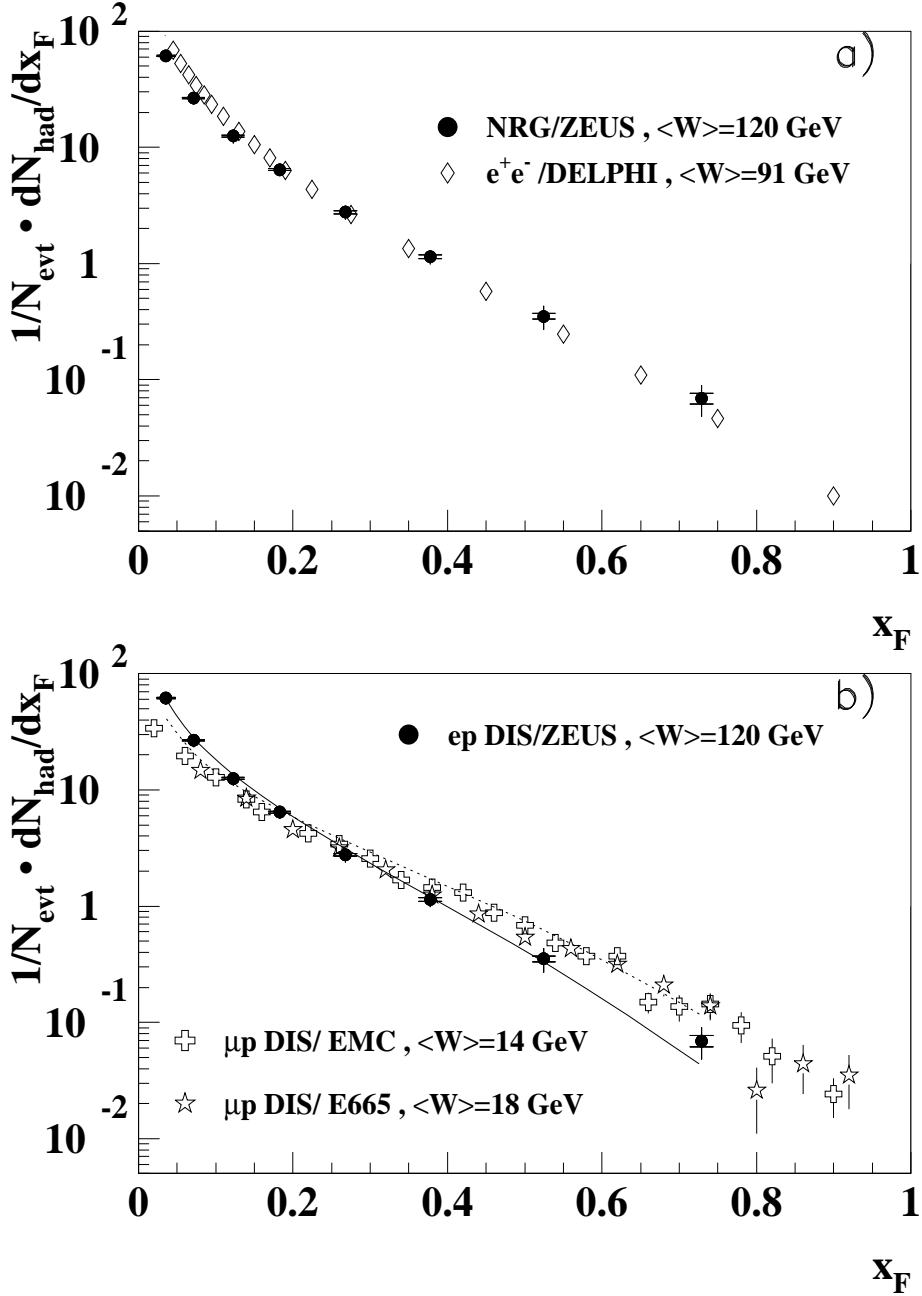}
           }
 \caption{
   a) $x_F$ distribution from this analysis (NRG events) compared to
   results from $e^+e^-$ annihilation on the $Z^0$ resonance ($W=91\gev$)
   \protect \cite{delphi91}.
   b) $x_F$ distribution from this analysis (NRG + LRG events) compared
   with results from $\mu p$ DIS  at $<W>=14\gev$ \protect \cite{EMC87b}
   and  at $<W>=18\gev$ \protect \cite{e665z}.
   In Fig.~6b) the solid curve shows the prediction
   of the MEPS model calculation and
   the dotted curve that of the QPM at HERA energies.
        }
 \label{pxfna9}
 \end{figure}

  In Fig.~\ref{pxfna9}a the $x_F$ distribution from the NRG events
  is compared with that from $e^+e^-$
  annihilation events on the $Z^0$ resonance \cite{delphi91}, where
  the value of the \epem\ centre-of-mass energy
  is comparable to the value of $W$ in the kinematic range analysed here.
  The differential rates for hadron production in $e^+e^-$
  annihilation were divided by two so that they correspond to a single
  hemisphere and can be directly compared with the results from DIS.
  The differential hadron multiplicity distribution  in DIS at HERA
  energies agrees with that observed in
  $e^+e^-$ collision events for $x_F \sgeq 0.1$.
  This  confirms the approximate independence
  of the hadron formation process from the
  type of the primary scattering objects, which most of the models assume
  \cite{string80,ff78}.

  The $x_F$ and  $<p_t^{*\,2}>$ distributions from this analysis  are compared
  with those of DIS events at lower values of $W$
  \cite{EMC87b,e665z}.
   Since in fixed target experiments the DIS event sample has not been
  separated into NRG and LRG events, the NRG and LRG event samples
  have been combined for the comparison.
   The $x_F$ and $p_t^*$ distribution as well as \ptm as a function of $x_F$
  for the NRG+LRG event sample are given in Table~\ref{distab}.
  The distributions have been corrected using a combination of
  Monte Carlo event samples
  generated by the POMPYT and the CDMBGF Monte Carlo generator.
  The relative normalisation of the Monte Carlo samples has
  been fixed by fitting the sum of the reconstructed $\eta_{max}$ distribution
  from the POMPYT and the CDMBGF Monte Carlo sample to
   the measured $\eta_{max}$ distribution \cite{lrgefl}.

  Figure~\ref{pxfna9}b shows that the $x_F$ distribution becomes
  significantly softer with increasing $W$.
  The prediction of the QPM, where no scale breaking effects
  due to QCD radiation are included,
  almost agrees with the result from the fixed target experiments
 \cite{EMC87a,EMC87b,e665z} but is very different from the result at
  HERA energies (dotted line in Fig.~\ref{pxfna9}b).
  The effects of  scaling violation 
  in the $x_F$ distributions of
  hadrons, which have been found to be small when measured in a
  limited interval of $W$ and $Q^2$ \cite{EMC82a,EMC91a},
  become evident when studied over a   large range of $W$ and $Q^2$.
  Models in which higher order $\alpha_s$ processes are considered (e.g.
  the MEPS model indicated by the full
  line in Fig.~\ref{pxfna9}b) agree reasonably   with the ZEUS data.

\begin{figure}
 \centerline{
 \ \hspace*{1.0cm}
 \epsfxsize=15cm
 \epsfysize=12cm
 \epsffile{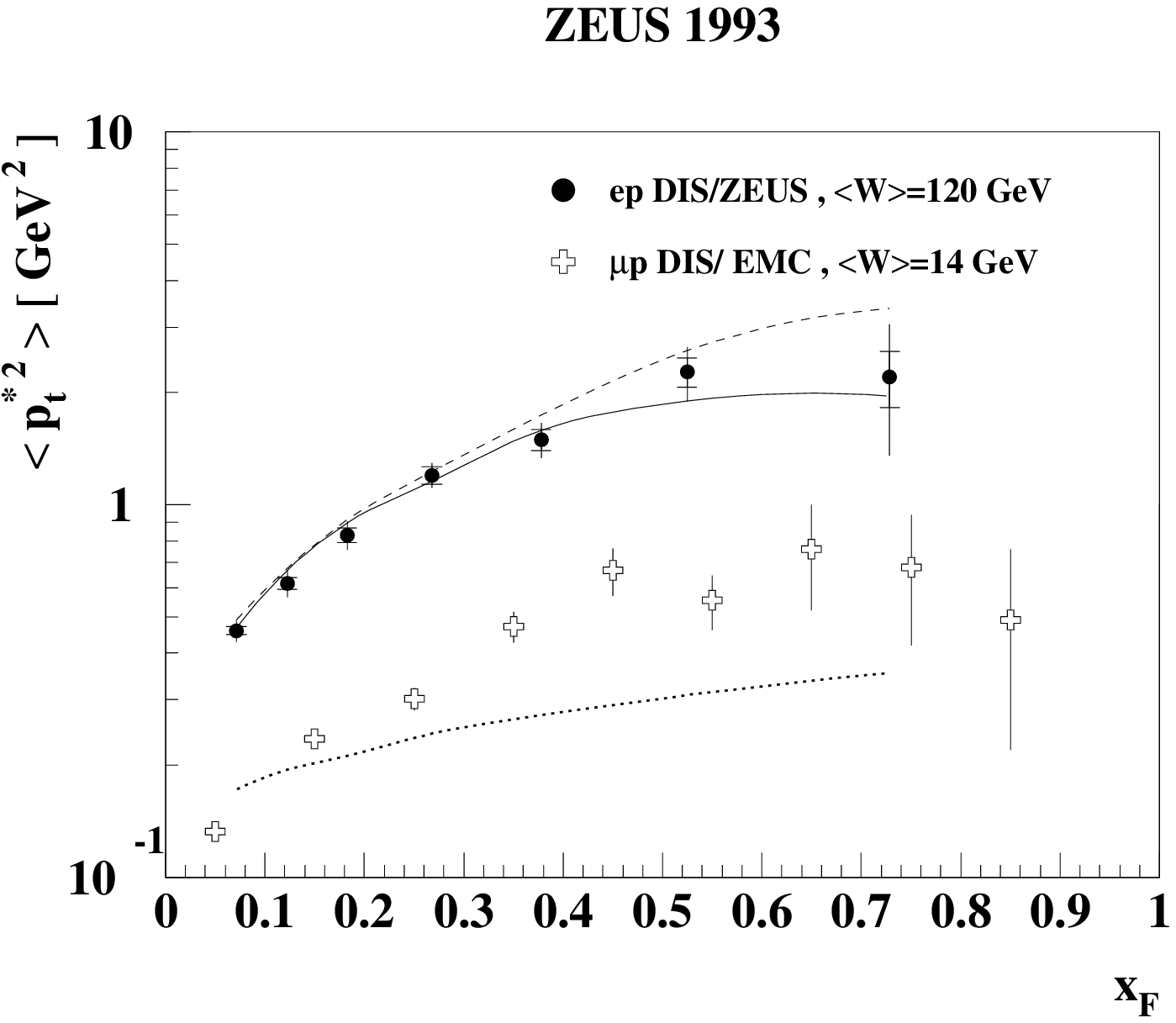}
            }
 \caption{
    $<\!p_t^{*\,2}\!>$ as a function of $x_F$ from this analysis
    (NRG + LRG events) compared to
    results from $\mu p$ DIS  at $<W>=14\gev$ \protect \cite{EMC87b}.
    The curves show results from model calculations at HERA energy with
    the MEPS model ({solid curve}), the CDMBGF model ({dashed curve})
    and the QPM ({dotted curve}).
         }
  \label{pptna9}
  \end{figure}

  The mean value of $p_t^{*\,2}$ as a function of $x_F$ is shown
  in Fig.~\ref{pptna9}
  for $<\!W\!> = 120 \gev$ (this analysis) and for $<\!W\!> = 14 \gev$
  from the EMC collaboration  \cite{EMC87b}.
  Comparing the results at low $W$ and high $W$ there
  is a strong increase of $<\!p_t^{*\,2}\!>$
  by a factor of about three over the whole range of $x_F>0.05$
  going from $W=14$ to 120~\gev .
  The comparison of the prediction from the QPM and the models including
  higher order QCD processes shows that
  QCD effects are much larger at HERA energies than
  at energies reached in fixed target experiments.

\begin{figure}[tp]
 \centerline{ \epsfxsize=16cm
              \epsfysize=10cm
              \epsffile{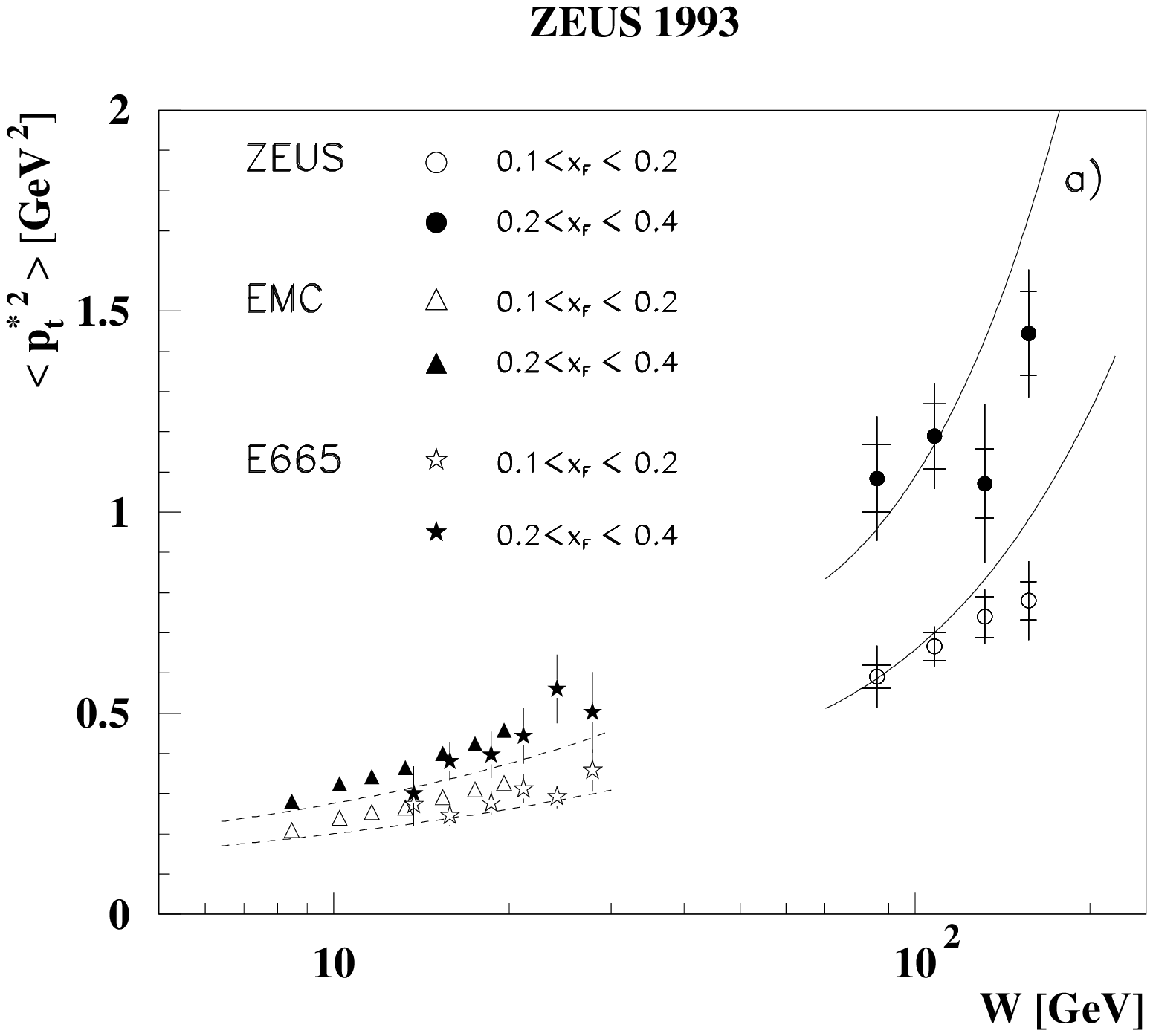}
            }
 \ \hspace*{0.3cm}
 \epsfxsize=16cm
 \epsfysize=8cm
 \epsffile{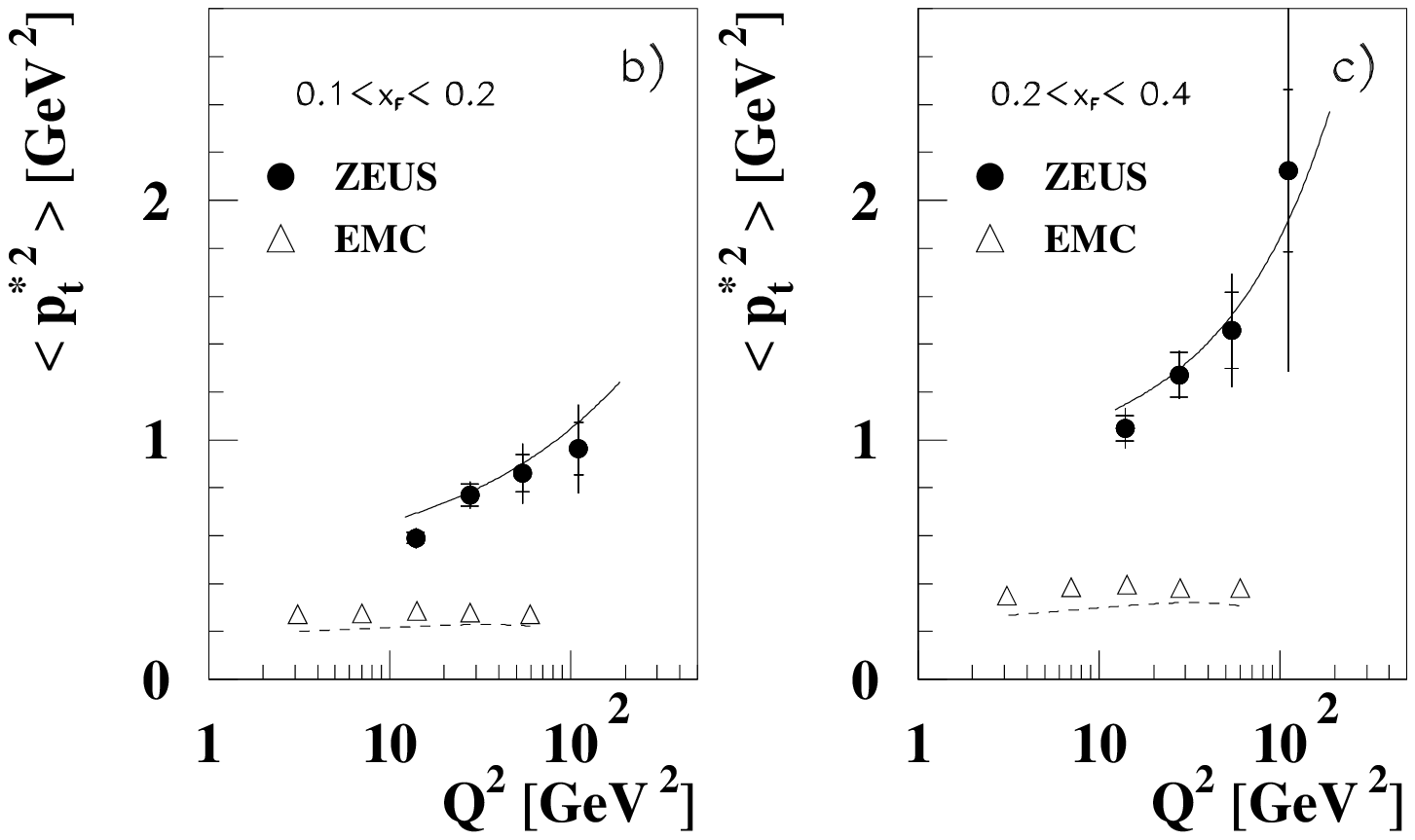}
 \caption{
     $<\!p_t^{*\,2}\!>$ in two intervals of $x_F$ as a function of
     a) $W$ and b),c) $Q^2$ compared with results from $\mu p$ DIS
     experiments (EMC \protect \cite{EMC91a} and E665 \protect
     \cite{e665pt}). The prediction of the MEPS Monte Carlo calculation
     is compared with the results of this analysis (solid curve) and of
     \protect \cite{EMC91a} (dashed curve).
         }
 \label{pptna2}
 \end{figure}

  For  a further analysis of the $W$ and $Q^2$ dependence,
  $<\!p_t^{*\,2}\!>$ was determined for two
  intervals in $x_F$ and four bins of $W$ at an average value for $Q^2$ of
  $28 \gevv$ (Fig.~\ref{pptna2}a)
  and four bins of $Q^2$ keeping $W$ fixed at an average
  value   of $120 \gev$ (Fig.~\ref{pptna2}b).
  The value of $<\!p_t^{*\,2}\!>$ increases both
  with $W$ and with $Q^2$. The results are tabulated in the
  Tables~\ref{wdeptab} and ~\ref{qdeptab}.

  These results are compared with those
  from a fixed target experiment  at lower energies \cite{EMC91a,e665pt}.
  The rise of  $<\!p_t^{*\,2}\!>$ with $W$, which had been observed already
  in the fixed target DIS experiments, continues in the range of $W$ seen
  at HERA.   However, the
  $Q^2$-dependence in these two ranges of $W$ is different. There is a large
  overlap of the $Q^2$ intervals covered. At HERA energies a rise of
  $<\!p_t^{*\,2}\!>$ with $Q^2$ is observed, while at low $W$ almost no
  dependence on $Q^2$ was found \cite{EMC91a}.

\begin{figure}[tp]
 \epsfxsize=18cm
 \epsfysize=10cm
 \epsffile{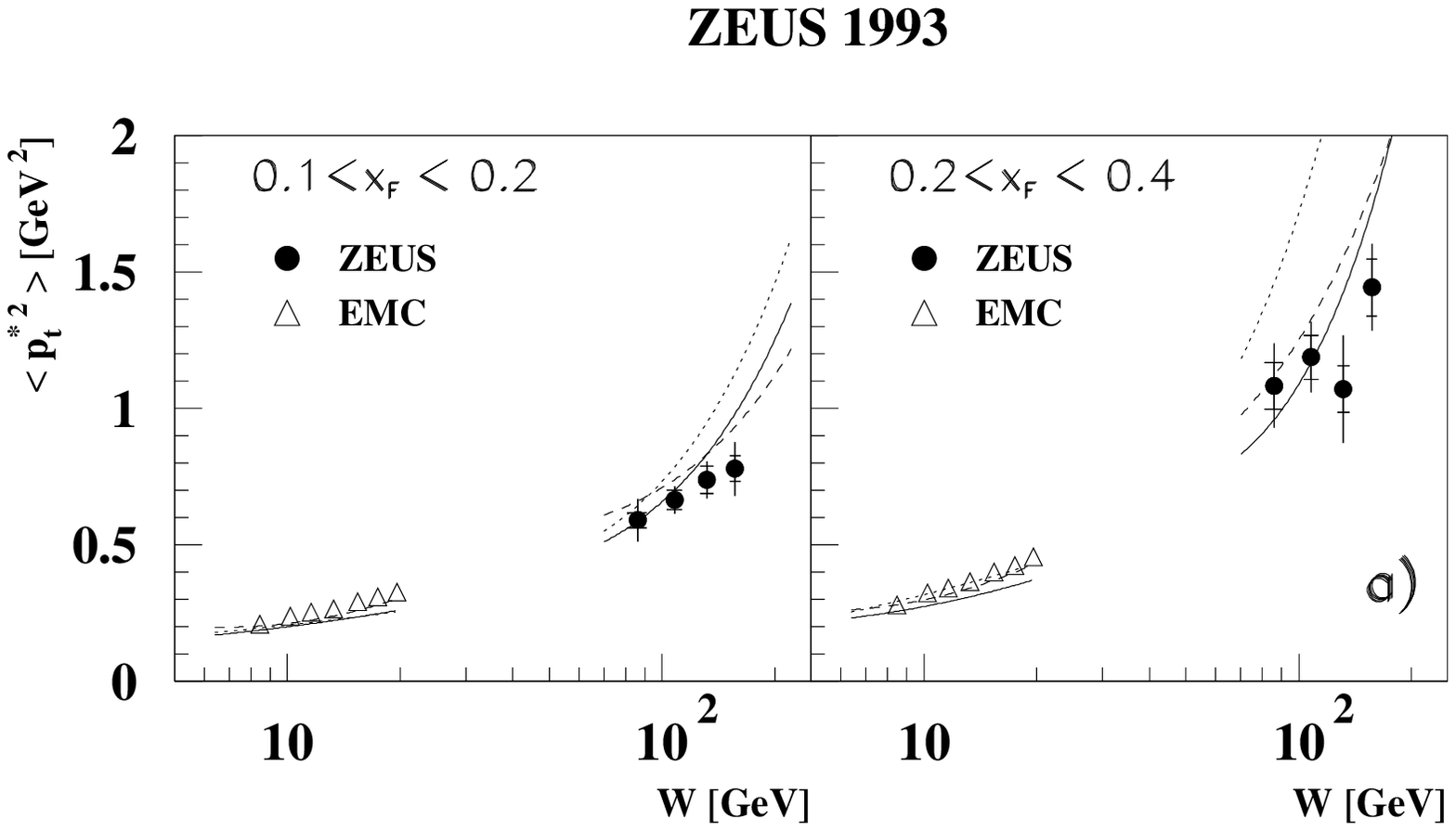}
 \epsfxsize=18cm
 \epsfysize=9cm
 \epsffile{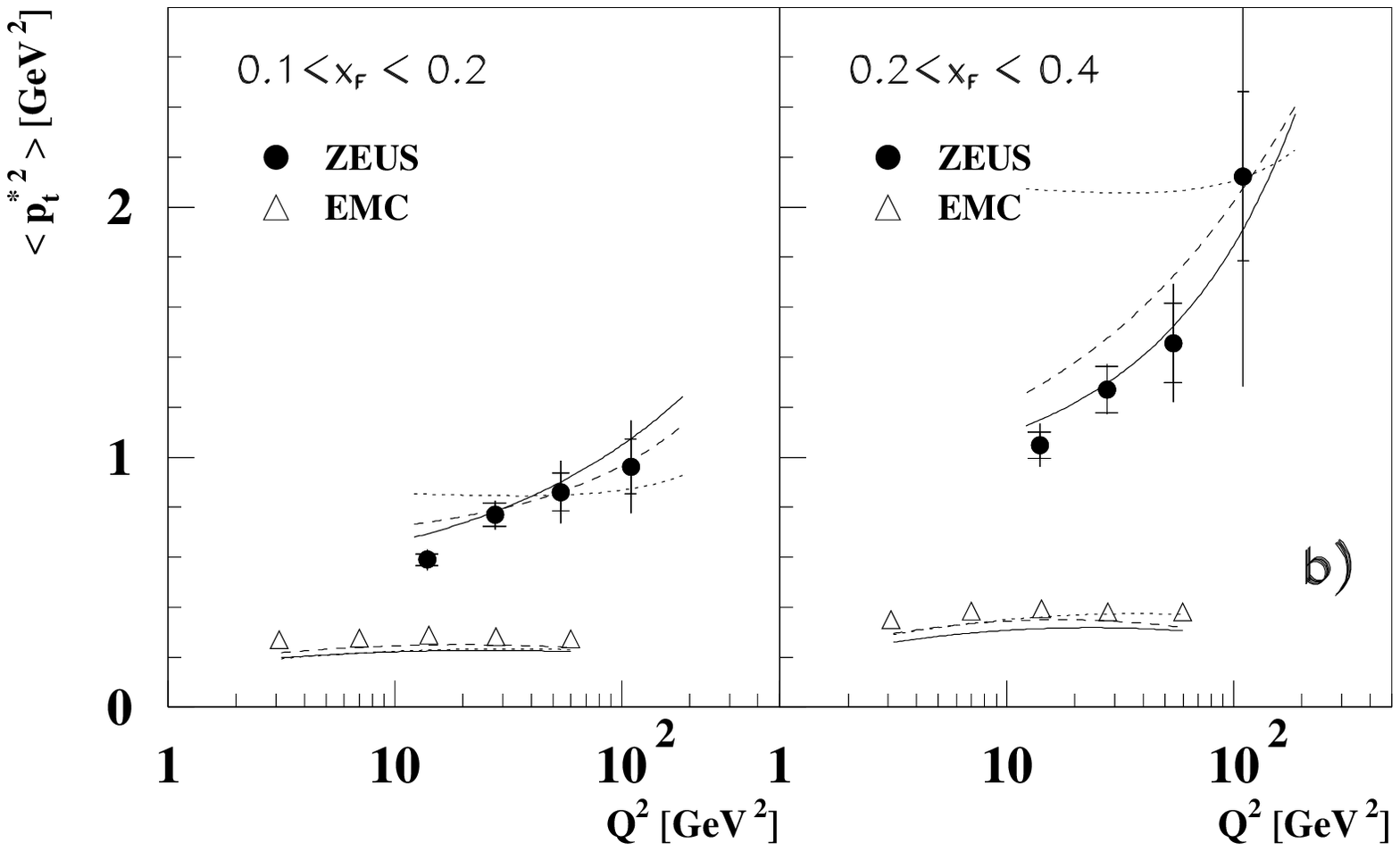}
 \caption{
       a) $W$ and b) $Q^2$ dependence of $<\! p_t^{*\,2}\!>$
       from EMC \protect \cite{EMC91a} and ZEUS data compared with
       different model predictions for the hadron formation:
       MEPS (solid curve), CDMBGF (dashed curve) and CDM (dotted curve).
       For the explanation of acronyms see Table~\protect\ref{mctab}.
         }
 \label{pwqzemc1}
 \end{figure}

  The results from the ZEUS experiment and the fixed target experiment are
  compared with model calculations in Fig.~\ref{pwqzemc1}.
  The $W$-dependence is reasonably described by the MEPS (solid line)
  and CDMBGF models (dashed line).
  Also the colour dipole model
  without including the BGF process (dotted line) qualitatively
  reproduces the $W$ dependence of $<\!p_t^{*\,2}\!>$ but overestimates the
  absolute value.
  The $Q^2$-dependence is also described by the MEPS and CDMBGF model but
  not by the colour dipole model (CDM) alone.
  The colour dipole model simulates higher order gluon radiation processes but
  the BGF process is not considered.
  The $Q^2$ dependence of \ptm  shows that it is necessary to include
  the explicit treatment of the BGF process as well in the simulation.

  \section{ Conclusions }

  Measurements of differential charged hadron multiplicity distributions in
  DIS events have been presented in the centre-of-mass system
  of the virtual photon and the proton
  at a centre-of-mass energy of $296 \gev$ for $10\le Q^2 \le 160 \gevv$
  and  $75\le W \le 175 \gev $.

  The transverse momentum, $p_t^*$, and $x_F$ distributions have been
  investigated separately for events with (LRG) and without a large
  rapidity gap (NRG) between the proton direction and the
  observed hadronic final state. In the whole range of $x_F>0.05$ the
  values of $< p_t^{*\,2}>$ for NRG events are much
  larger than  those for the LRG events.
  These results confirm that gluon radiation in LRG events is strongly
  suppressed as compared to `standard' DIS events at comparable $W$.
  A comparison of the data with the prediction of the QPM shows, however,
  that some QCD radiation is present also in LRG events.

  The value of $<\!p_t^{*\,2}\!>$ in the LRG events is similar to that
  observed in deep inelastic $\mu p$ scattering experiments on fixed targets
  at low $W$ ($<\!W\!> =14 \gev$). This indicates that the
  multi-particle production in LRG events is similar to that in DIS at
  a scale of the final state invariant mass $W=M_X$,
  where $M_X$ is the invariant mass of the observed hadronic final state
  $X$, excluding the proton.

  The comparison of the $x_F$ distributions in $e^+e^-$ annihilation and
  in DIS events confirms
  the hypothesis that the hadron formation process in the current jet
  region is  approximately independent of the type
  of the primary interacting particles.

  The comparison of results presented here
  with those of DIS at low $W$ from fixed target experiments
  allows a study of the development of QCD effects in
  the $x_F$ and  $p_t^{*}$ distributions over a large range in $W$ and $Q^2$.
  A significant increase of $<\!p_t^{*\,2}\!>$ with $W$ is found.
  At HERA energies, the   mean value of $p_t^{*\,2}$
  also rises with increasing $Q^2$ at fixed $W$.
  This can be understood in terms of the increase of the momentum
  space allowing the formation of more multi-jet events.

\section*{Acknowledgements}
The experiment was made possible by the inventiveness and the diligent
efforts of the HERA machine group who continued to run HERA most
efficiently during 1993.

The design, construction and installation of the ZEUS detector has
been made possible by the ingenuity and dedicated effort of many people
from inside DESY and from the home institutes, who are not listed as authors.
Their contributions are acknowledged with great appreciation.

The strong support and encouragement of the DESY Directorate
has been invaluable.

We also gratefully acknowledge the support of the DESY computing and network
services.


\begin{table}[tp]
\centering
\begin{tabular}{|c|c|c|}
\hline
\multicolumn{3}{|c|}{ {\bf A} } \\
\hline
 $x_F$  & $<\!x_F\!>$ & $\frac{1}{N_{evt}}\frac{dN_{had}}{dx_{F}}$      \\
\hline
    0.03 \,-\,    0.05 &     0.04 &
       64.96 \hspace*{0.1cm} $\pm$   0.78 \hspace*{0.1cm} $\pm$  10.53 \\
    0.05 \,-\,    0.10 &     0.07 &
       27.92 \hspace*{0.1cm} $\pm$   0.34 \hspace*{0.1cm} $\pm$   4.27 \\
    0.10 \,-\,    0.15 &     0.12 &
       12.89 \hspace*{0.1cm} $\pm$   0.23 \hspace*{0.1cm} $\pm$   1.97 \\
    0.15 \,-\,    0.22 &     0.18 &
        6.67 \hspace*{0.1cm} $\pm$   0.13 \hspace*{0.1cm} $\pm$   0.97 \\
    0.22 \,-\,    0.32 &     0.27 &
        2.86 \hspace*{0.1cm} $\pm$   0.07 \hspace*{0.1cm} $\pm$   0.45 \\
    0.32 \,-\,    0.45 &     0.38 &
        1.15 \hspace*{0.1cm} $\pm$   0.04 \hspace*{0.1cm} $\pm$   0.15 \\
    0.45 \,-\,    0.65 &     0.52 &
        0.36 \hspace*{0.1cm} $\pm$   0.02 \hspace*{0.1cm} $\pm$   0.08 \\
    0.65 \,-\,    0.90 &     0.73 &
        0.07 \hspace*{0.1cm} $\pm$   0.006 \hspace*{0.1cm} $\pm$   0.02 \\
\hline
\multicolumn{3}{|c|}{ {\bf B} } \\
\hline
 $p_t^*$ $\gev$  & $<\!p_t^*\!>$ $\gev$ &
   $\frac{1}{N_{evt}}\frac{dN_{had}}{dp_t^*}$ $\gev^{-1} $    \\
\hline
    0.00 \,-\,    0.10 &     0.07 &
        1.05 \hspace*{0.1cm} $\pm$   0.05 \hspace*{0.1cm} $\pm$   0.18 \\
    0.10 \,-\,    0.20 &     0.15 &
        2.84 \hspace*{0.1cm} $\pm$   0.08 \hspace*{0.1cm} $\pm$   0.51 \\
    0.20 \,-\,    0.40 &     0.30 &
        3.89 \hspace*{0.1cm} $\pm$   0.07 \hspace*{0.1cm} $\pm$   0.60 \\
    0.40 \,-\,    0.60 &     0.49 &
        3.34 \hspace*{0.1cm} $\pm$   0.06 \hspace*{0.1cm} $\pm$   0.47 \\
    0.60 \,-\,    0.80 &     0.69 &
        2.24 \hspace*{0.1cm} $\pm$   0.05 \hspace*{0.1cm} $\pm$   0.33 \\
    0.80 \,-\,    1.20 &     0.96 &
        1.10 \hspace*{0.1cm} $\pm$   0.02 \hspace*{0.1cm} $\pm$   0.18 \\
    1.20 \,-\,    1.50 &     1.33 &
        0.47 \hspace*{0.1cm} $\pm$   0.02 \hspace*{0.1cm} $\pm$   0.07 \\
    1.50 \,-\,    2.00 &     1.71 &
        0.23 \hspace*{0.1cm} $\pm$   0.01 \hspace*{0.1cm} $\pm$   0.03 \\
    2.00 \,-\,    2.75 &     2.30 &
        0.08 \hspace*{0.1cm} $\pm$   0.004 \hspace*{0.1cm} $\pm$   0.01 \\
    2.75 \,-\,    3.50 &     3.07 &
        0.03 \hspace*{0.1cm} $\pm$   0.002 \hspace*{0.1cm} $\pm$   0.004 \\
    3.50 \,-\,    5.00 &     4.04 &
        0.01 \hspace*{0.1cm} $\pm$   0.001 \hspace*{0.1cm} $\pm$   0.001 \\
\hline
\multicolumn{3}{|c|}{ {\bf C} } \\
\hline
 $x_F$ & $<\!x_F\!>$ & $<\!p_t^{*\,2}\!>$ $ \gevv $ \\
\hline
    0.05 \,-\,    0.10 &     0.07 &
        0.47 \hspace*{0.1cm} $\pm$   0.01 \hspace*{0.1cm} $\pm$   0.02 \\
    0.10 \,-\,    0.15 &     0.12 &
        0.63 \hspace*{0.1cm} $\pm$   0.02 \hspace*{0.1cm} $\pm$   0.04 \\
    0.15 \,-\,    0.22 &     0.18 &
        0.85 \hspace*{0.1cm} $\pm$   0.04 \hspace*{0.1cm} $\pm$   0.03 \\
    0.22 \,-\,    0.32 &     0.27 &
        1.19 \hspace*{0.1cm} $\pm$   0.06 \hspace*{0.1cm} $\pm$   0.06 \\
    0.32 \,-\,    0.45 &     0.38 &
        1.50 \hspace*{0.1cm} $\pm$   0.09 \hspace*{0.1cm} $\pm$   0.13 \\
    0.45 \,-\,    0.65 &     0.52 &
        2.30 \hspace*{0.1cm} $\pm$   0.19 \hspace*{0.1cm} $\pm$   0.33 \\
    0.65 \,-\,    0.90 &     0.73 &
        2.09 \hspace*{0.1cm} $\pm$   0.33 \hspace*{0.1cm} $\pm$   0.66 \\
\hline
\end{tabular}
\caption{ Differential multiplicities for charged hadrons as a function
 of A) $x_F$ and B) $p_t^{*}$ ($x_F>0.05$)
    and C)  $<\!p_t^{*\,2}\!>$ as a function of $x_F$
  for DIS events with $\eta_{max}>1.5$ (NRG)
  in the range of $10<Q^2<160 \gevv$ and $75<W<175 \gev$.
   Statistical and systematic errors are given separately.
}
\label{nrgtab}
\end{table}

\begin{table}[tp]
\centering
\begin{tabular}{|c|c|c|}
\hline
\multicolumn{3}{|c|}{ {\bf A} } \\
\hline
 $x_F$ & $<\!x_F\!>$ & $\frac{1}{N_{evt}}\frac{dN_{had}}{dx_{F}}$ \\
\hline
    0.03 \,-\,    0.05 &     0.04 &
       26.97 \hspace*{0.1cm} $\pm$   1.88 \hspace*{0.1cm} $\pm$   2.55 \\
    0.05 \,-\,    0.10 &     0.07 &
       19.70 \hspace*{0.1cm} $\pm$   1.24 \hspace*{0.1cm} $\pm$   0.94 \\
    0.10 \,-\,    0.15 &     0.12 &
       12.16 \hspace*{0.1cm} $\pm$   1.02 \hspace*{0.1cm} $\pm$   1.51 \\
    0.15 \,-\,    0.22 &     0.18 &
        6.49 \hspace*{0.1cm} $\pm$   0.59 \hspace*{0.1cm} $\pm$   1.01 \\
    0.22 \,-\,    0.32 &     0.27 &
        4.02 \hspace*{0.1cm} $\pm$   0.45 \hspace*{0.1cm} $\pm$   0.82 \\
    0.32 \,-\,    0.45 &     0.38 &
        1.41 \hspace*{0.1cm} $\pm$   0.20 \hspace*{0.1cm} $\pm$   0.46 \\
    0.45 \,-\,    0.65 &     0.54 &
        0.68 \hspace*{0.1cm} $\pm$   0.12 \hspace*{0.1cm} $\pm$   0.10 \\
    0.65 \,-\,    0.90 &     0.75 &
        0.31 \hspace*{0.1cm} $\pm$   0.07 \hspace*{0.1cm} $\pm$   0.07 \\
\hline
\multicolumn{3}{|c|}{ {\bf B} } \\
\hline
 $p_t^*$ $\gev$  & $<\!p_t^*\!>$ $\gev$ &
   $\frac{1}{N_{evt}}\frac{dN_{had}}{dp_t^*}$ $\gev^{-1} $    \\
\hline
    0.00 \,-\,    0.10 &     0.07 &
        1.36 \hspace*{0.1cm} $\pm$   0.05 \hspace*{0.1cm} $\pm$   0.50 \\
    0.10 \,-\,    0.20 &     0.15 &
        4.31 \hspace*{0.1cm} $\pm$   0.11 \hspace*{0.1cm} $\pm$   1.11 \\
    0.20 \,-\,    0.40 &     0.30 &
        4.91 \hspace*{0.1cm} $\pm$   0.05 \hspace*{0.1cm} $\pm$   0.41 \\
    0.40 \,-\,    0.60 &     0.49 &
        3.44 \hspace*{0.1cm} $\pm$   0.04 \hspace*{0.1cm} $\pm$   0.34 \\
    0.60 \,-\,    0.80 &     0.68 &
        1.45 \hspace*{0.1cm} $\pm$   0.02 \hspace*{0.1cm} $\pm$   0.55 \\
    0.80 \,-\,    1.20 &     0.93 &
        0.64 \hspace*{0.1cm} $\pm$   0.01 \hspace*{0.1cm} $\pm$   0.045 \\
    1.20 \,-\,    2.00 &     1.41 &
        0.09 \hspace*{0.1cm} $\pm$   0.003 \hspace*{0.1cm} $\pm$   0.02 \\
    2.00 \,-\,    5.00 &     3.86 &
        0.001 \hspace*{0.1cm} $\pm$   0.001 \hspace*{0.1cm} $\pm$   0.004 \\
\hline
\multicolumn{3}{|c|}{ {\bf C} } \\
\hline
 $x_F$ & $<\!x_F\!>$ & $<\!p_t^{*\,2}\!> $ $ \gevv $ \\
\hline
    0.05 \,-\,    0.10 &     0.07 &
        0.19 \hspace*{0.1cm} $\pm$   0.02 \hspace*{0.1cm} $\pm$   0.01 \\
    0.10 \,-\,    0.15 &     0.12 &
        0.24 \hspace*{0.1cm} $\pm$   0.05 \hspace*{0.1cm} $\pm$   0.03 \\
    0.15 \,-\,    0.22 &     0.18 &
        0.35 \hspace*{0.1cm} $\pm$   0.08 \hspace*{0.1cm} $\pm$   0.04 \\
    0.22 \,-\,    0.32 &     0.27 &
        0.41 \hspace*{0.1cm} $\pm$   0.11 \hspace*{0.1cm} $\pm$   0.16 \\
    0.32 \,-\,    0.45 &     0.38 &
        0.38 \hspace*{0.1cm} $\pm$   0.07 \hspace*{0.1cm} $\pm$   0.11 \\
    0.45 \,-\,    0.65 &     0.54 &
        0.50 \hspace*{0.1cm} $\pm$   0.10 \hspace*{0.1cm} $\pm$   0.12 \\
    0.65 \,-\,    0.90 &     0.75 &
        0.37 \hspace*{0.1cm} $\pm$   0.13 \hspace*{0.1cm} $\pm$   0.60 \\
\hline
\end{tabular}
\caption{ Differential multiplicities for charged hadrons as a function
 of A) $x_F$ and B) $p_t^{*}$ ($x_F>0.05$)
    and  C) $<\!p_t^{*\,2}\!>$ as a function of $x_F$
  for DIS events with $\eta_{max}<1.5$ (LRG)
  in the range of $10<Q^2<160 \gevv$ and $75<W<175 \gev$.
   Statistical and systematic errors are given separately.
   }
\label{lrgtab}
\end{table}

\begin{table}[tp]
\centering
\begin{tabular}{|c|c|c|}
\hline
\multicolumn{3}{|c|}{ {\bf A} } \\
\hline
 $x_F$  & $<\!x_F\!>$ & $\frac{1}{N_{evt}}\frac{dN_{had}}{dx_{F}}$      \\
\hline
    0.03 \,-\,    0.05 &     0.04 &
       62.99 \hspace*{0.1cm} $\pm$   0.75 \hspace*{0.1cm} $\pm$   9.59 \\
    0.05 \,-\,    0.10 &     0.07 &
       27.48 \hspace*{0.1cm} $\pm$   0.33 \hspace*{0.1cm} $\pm$   4.03 \\
    0.10 \,-\,    0.15 &     0.12 &
       12.86 \hspace*{0.1cm} $\pm$   0.22 \hspace*{0.1cm} $\pm$   1.99 \\
    0.15 \,-\,    0.22 &     0.18 &
        6.67 \hspace*{0.1cm} $\pm$   0.13 \hspace*{0.1cm} $\pm$   0.98 \\
    0.22 \,-\,    0.32 &     0.27 &
        2.90 \hspace*{0.1cm} $\pm$   0.07 \hspace*{0.1cm} $\pm$   0.46 \\
    0.32 \,-\,    0.45 &     0.38 &
        1.18 \hspace*{0.1cm} $\pm$   0.04 \hspace*{0.1cm} $\pm$   0.17 \\
    0.45 \,-\,    0.65 &     0.52 &
        0.37 \hspace*{0.1cm} $\pm$   0.02 \hspace*{0.1cm} $\pm$   0.08 \\
    0.65 \,-\,    0.90 &     0.73 &
        0.08 \hspace*{0.1cm} $\pm$   0.007 \hspace*{0.1cm} $\pm$   0.03 \\
\hline
\multicolumn{3}{|c|}{ {\bf B} } \\
\hline
 $p_t^*$ $\gev$  & $<\!p_t^*\!>$ $\gev$ &
   $\frac{1}{N_{evt}}\frac{dN_{had}}{dp_t^*}$ $\gev^{-1} $    \\
\hline
    0.00 \,-\,    0.10 &     0.07 &
        1.10 \hspace*{0.1cm} $\pm$   0.05 \hspace*{0.1cm} $\pm$   0.18 \\
    0.10 \,-\,    0.20 &     0.15 &
        2.92 \hspace*{0.1cm} $\pm$   0.08 \hspace*{0.1cm} $\pm$   0.52 \\
    0.20 \,-\,    0.40 &     0.30 &
        3.98 \hspace*{0.1cm} $\pm$   0.07 \hspace*{0.1cm} $\pm$   0.60 \\
    0.40 \,-\,    0.60 &     0.49 &
        3.35 \hspace*{0.1cm} $\pm$   0.06 \hspace*{0.1cm} $\pm$   0.47 \\
    0.60 \,-\,    0.80 &     0.69 &
        2.22 \hspace*{0.1cm} $\pm$   0.05 \hspace*{0.1cm} $\pm$   0.32 \\
    0.80 \,-\,    1.20 &     0.96 &
        1.07 \hspace*{0.1cm} $\pm$   0.02 \hspace*{0.1cm} $\pm$   0.18 \\
    1.20 \,-\,    1.50 &     1.33 &
        0.45 \hspace*{0.1cm} $\pm$   0.02 \hspace*{0.1cm} $\pm$   0.08 \\
    1.50 \,-\,    2.00 &     1.71 &
        0.22 \hspace*{0.1cm} $\pm$   0.01 \hspace*{0.1cm} $\pm$   0.03 \\
    2.00 \,-\,    2.75 &     2.30 &
        0.07 \hspace*{0.1cm} $\pm$   0.004 \hspace*{0.1cm} $\pm$   0.01 \\
    2.75 \,-\,    3.50 &     3.07 &
        0.02 \hspace*{0.1cm} $\pm$   0.002 \hspace*{0.1cm} $\pm$   0.004 \\
    3.50 \,-\,    5.00 &     4.04 &
        0.01 \hspace*{0.1cm} $\pm$   0.001 \hspace*{0.1cm} $\pm$   0.001 \\
\hline
\multicolumn{3}{|c|}{ {\bf C} } \\
\hline
 $x_F$ & $<\!x_F\!>$ & $<\!p_t^{*\,2}\!>$ $ \gevv $ \\
\hline
    0.05 \,-\,    0.10 &     0.07 &
        0.46 \hspace*{0.1cm} $\pm$   0.01 \hspace*{0.1cm} $\pm$   0.02 \\
    0.10 \,-\,    0.15 &     0.12 &
        0.61 \hspace*{0.1cm} $\pm$   0.02 \hspace*{0.1cm} $\pm$   0.04 \\
    0.15 \,-\,    0.22 &     0.18 &
        0.82 \hspace*{0.1cm} $\pm$   0.04 \hspace*{0.1cm} $\pm$   0.03 \\
    0.22 \,-\,    0.32 &     0.27 &
        1.14 \hspace*{0.1cm} $\pm$   0.06 \hspace*{0.1cm} $\pm$   0.06 \\
    0.32 \,-\,    0.45 &     0.38 &
        1.40 \hspace*{0.1cm} $\pm$   0.08 \hspace*{0.1cm} $\pm$   0.11 \\
    0.45 \,-\,    0.65 &     0.52 &
        2.13 \hspace*{0.1cm} $\pm$   0.18 \hspace*{0.1cm} $\pm$   0.39 \\
    0.65 \,-\,    0.90 &     0.73 &
        1.82 \hspace*{0.1cm} $\pm$   0.28 \hspace*{0.1cm} $\pm$   0.56 \\
\hline
\end{tabular}
\caption{ Differential multiplicities for charged hadrons as a function
 of A) $x_F$ and B) $p_t^{*}$ ($x_F>0.05$)
    and C)  $<\!p_t^{*\,2}\!>$ as a function of $x_F$
  for DIS events (combined NRG + LRG event sample)
  in the range of $10<Q^2<160 \gevv$ and $75<W<175 \gev$.
   Statistical and systematic errors are given separately.
}
\label{distab}
\end{table}

\newpage

\begin{table}[h]
\centering
\begin{tabular}{|c|c|c|c|}
\hline
\multicolumn{2}{|c|}{ } &
\multicolumn{1}{|c|}{ $ 0.1 < x_F <0.2$ } &
\multicolumn{1}{|c|}{ $ 0.2 < x_F <0.4$ } \\
\hline
 $W  $ $\gev $ & $<\!W\!>$  $\gev$ & $<\!p_t^{*\,2}\!> $ $ \gevv $  &
                       $<\!p_t^{*\,2}\!> $ $ \gevv $  \\
\hline
   77 \,-\,   95 &    86 &
         0.59 \hspace*{0.1cm} $\pm$    0.03 \hspace*{0.1cm} $\pm$    0.07 &
         1.08 \hspace*{0.1cm} $\pm$    0.08 \hspace*{0.1cm} $\pm$    0.13 \\
   95 \,-\,  122 &   108 &
         0.67 \hspace*{0.1cm} $\pm$    0.03 \hspace*{0.1cm} $\pm$    0.04 &
         1.19 \hspace*{0.1cm} $\pm$    0.08 \hspace*{0.1cm} $\pm$    0.10 \\
  122 \,-\,  141 &   132 &
         0.74 \hspace*{0.1cm} $\pm$    0.05 \hspace*{0.1cm} $\pm$    0.04 &
         1.07 \hspace*{0.1cm} $\pm$    0.09 \hspace*{0.1cm} $\pm$    0.18 \\
  141 \,-\,  173 &   157 &
         0.78 \hspace*{0.1cm} $\pm$    0.05 \hspace*{0.1cm} $\pm$    0.09 &
         1.44 \hspace*{0.1cm} $\pm$    0.10 \hspace*{0.1cm} $\pm$    0.12 \\
\hline
\end{tabular}
\caption{  $<\!p_t^{*\,2}\!>$ as a function of $W$ in two intervals of $x_F$.
   Statistical and systematic errors are given separately.
   }
\label{wdeptab}
\end{table}

\begin{table}[h]
\centering
\begin{tabular}{|c|c|c|c|}
\hline
\multicolumn{2}{|c|}{ } &
\multicolumn{1}{|c|}{ $ 0.1 < x_F <0.2$ } &
\multicolumn{1}{|c|}{ $ 0.2 < x_F <0.4$ } \\
\hline
 $Q^2 $ $\gevv$  & $<\!Q^2\!>$ $ \gevv$ & $<\!p_t^{*\,2}\!> $ $\gevv $  &
                      $<\!p_t^{*\,2}\!> $ $ \gevv $ \\
\hline
   10 \,-\,   20 &    14 &
         0.59 \hspace*{0.1cm} $\pm$    0.02 \hspace*{0.1cm} $\pm$    0.03 &
         1.05 \hspace*{0.1cm} $\pm$    0.05 \hspace*{0.1cm} $\pm$    0.07  \\
   20 \,-\,   40 &    28  &
         0.77 \hspace*{0.1cm} $\pm$    0.05 \hspace*{0.1cm} $\pm$    0.03 &
         1.27 \hspace*{0.1cm} $\pm$    0.09 \hspace*{0.1cm} $\pm$    0.04  \\
   40 \,-\,   80 &    54 &
         0.86 \hspace*{0.1cm} $\pm$    0.08 \hspace*{0.1cm} $\pm$    0.10 &
         1.46 \hspace*{0.1cm} $\pm$    0.16 \hspace*{0.1cm} $\pm$    0.17 \\
   80 \,-\,  160 &   110 &
         0.96 \hspace*{0.1cm} $\pm$    0.11 \hspace*{0.1cm} $\pm$    0.15 &
         2.12 \hspace*{0.1cm} $\pm$    0.34 \hspace*{0.1cm} $\pm$    0.77 \\
\hline
\end{tabular}
\caption{  $<\!p_t^{*\,2}\!>$ as a function of $Q^2$ in two intervals of $x_F$.
   Statistical and systematic errors are given separately.
   }
\label{qdeptab}
\end{table}


\end{document}